\documentclass[hyper]{JHEP3}
\usepackage{epsfig,axodraw,graphicx,cite,amsmath,psfrag}
\newcommand{\one}{{({\bf 1})} }
\newcommand{\met}{\rlap{\,/}E}
\newcommand{\oo}{{(1,1)} }

\newcommand{\gev}{\mbox{\ GeV}}
\newcommand{\SLASH}[2]{\makebox[#2ex][l]{$#1$}/}
\newcommand{\Eslash}{\SLASH{E}{.5}\,}

\title{\Large Two universal extra dimensions and spinless photons at the ILC { $\; $ } \\ }

\author{Ayres Freitas$^{1,2}$, Kyoungchul Kong$^{\,3}$ \\ \\ 

$^1$University of Chicago, 5640 S. Ellis Ave., chicago, IL 60637, USA   \\ \\
$^2$HEP Division, Argonne National Laboratory, 9700 Cass Ave., Argonne, IL 60439, USA   \\  \\
$^3$Theoretical Physics Department, Fermilab, Batavia, IL 60510, USA \\ \\
\email{afreitas@hep.anl.gov, kckong@fnal.gov} \\ 
}

\abstract{  We study the ILC phenomenology of Kaluza-Klein (KK) modes
along two universal extra dimensions  compactified on the chiral square.  We
compute production cross sections of various (1,0) particles at the ILC with
$\sqrt{s}=1$ TeV,  focusing on decays of KK-leptons and the KK partner of the
hypercharge gauge boson down to   the ``spinless photon'', which is the lightest
KK particle.  We contrast this model to one universal extra dimension with
KK-photon (spin-1) and  supersymmetry with neutralino (spin-1/2) or
gravitino (spin-3/2) dark matter. We also investigate the discovery potential
for (1,1) KK bosons as s-channel resonances.}

\preprint{{\small ANL-HEP-PR-07-99 }\\{\small  FERMILAB-PUB-07-619-T} \\
November 27, 2007 \\ } 

\begin{document}


\section{Introduction} 
\label{sec:introduction} 

Among many models for new physics beyond the standard model (SM), supersymmetry
(SUSY) and models with extra dimensions play a special role, since they
introduce a new symmetry which is felt by all particles. Therefore they lead to
a multiplication of the particle content with respect to the SM. The new
particles can be protected by a discrete symmetry, thus allowing for a natural
stable dark matter candidate, which usually is partner of the SM gauge bosons.
In the most widely known examples of the Minimal Supersymmetric standard model
(MSSM) and the standard model with one universal extra dimension (5DSM)
\cite{Appelquist:2000nn} the typical dark matter candidates are fermions
(neutralino or gravitino) or Kaluza-Klein (KK) vector bosons, 
respectively\footnote{It should be pointed out, however, that if the MSSM is
extended by a right-handed sneutrino superfield, the right-chiral sneutrino could
be a viable scalar dark matter candidate.}. In the standard model with two
universal extra dimensions (6DSM) \cite{Dobrescu:2004zi,Burdman:2005sr}, on the
other hand, the lightest Kaluza-Klein particle (LKP) is expected to be a scalar,
which is an adjoint under the gauge groups. It originates from the polarization
modes of the KK vector bosons along the extra dimensions, which upon
compactification receives a smaller mass than the usual four polarization modes
of the KK vector boson.

The structure of the 6DSM and its phenomenology at hadron colliders has been
studied in several articles \cite{Dobrescu:2004zi,Burdman:2005sr,
Ponton:2005kx,Burdman:2006gy,Dobrescu:2007xf,Dobrescu:2007yp}. Furthermore it
was shown that the scalar adjoint could provide a good dark matter candidate
\cite{Dobrescu:2007ec,Hsieh:2006qe,Hsieh:2006fg}. Although there might be some
tension between the region consistent with dark matter and the collider bounds
from Tevatron, a light LKP (relatively large radius $R$ of the extra dimensions)
is preferred. As a consequence, this is an excellent opportunity for an $e^+e^-$
linear collider like ILC with $\sqrt{s}=1$ TeV to cover much of the particle
spectrum.

If new particles are found that are compatible with supersymmetry or
extra-dimensional models, the determination of the spin of the stable dark
matter candidate will be crucial for understanding the experimental discoveries.
Since this stable particle is weakly interacting and escapes the detector
leaving a missing energy signal, it is very difficult to measure its spin at
hadron colliders, especially if the particle spectrum is nearly degenerate
 (see Refs.~\cite{Hooper:2007qk,Yavin:2007zz} and references therein).
However, with the advantage of a well-defined initial state
with controllable center-of-mass energy and beam polarization, spin
determination of new particles is possible without serious model assumptions.

In this paper, we study the phenomenology of KK-particles in the 6DSM at the
ILC with $\sqrt{s}=1$ TeV. In particular we analyze how the nature of the LKP
can be explored in the decay of other KK partners. Another interesting feature
of the 6DSM is the existence of modes with even KK parity and a mass that is
only larger by a factor of $\sqrt{2}$ with respect to the lowest KK modes---in
contrast to the 5DSM, where the KK mode masses are separated by integer ratios.
These particles can be produced singly at the ILC, so that they are quite likely
to be within the energy reach of the collider.

We start by reviewing the standard model in two universal extra dimensions in
section~\ref{sec:model} and proceed in section~\ref{sec:prod_decays} to 
calculate production cross sections of various (1,0) particles at an $e^+e^-$
linear collider with  $\sqrt{s} = 1$ TeV. We also discuss how those produced
particles decay, leading to missing energy signatures. In
section~\ref{sec:signatures} we then analyze how the properties of KK particles
can be determined, and point out methods to
discriminate the 6DSM  from 5DSM and supersymmetry. Section
\ref{sec:conclusions} is reserved for summary and discussion.

\section{The six dimensional standard model}\label{sec:model}

Universal extra dimensions (UED), in which all standard model fields
propagate,  have generated much interest due to opportunities for their
discovery  at collider and dark matter experiments. Precision electroweak
constraints do not place very stringent constraints on the compactification
scale of  universal dimensions, allowing for new particles that are light 
enough to be accessible at the current generation of hadron  colliders
\cite{Appelquist:2000nn} (see \cite{Hooper:2007qk} for review). Theories
with universal extra dimensions have a $Z_2$ symmetry  which is a remnant of
the higher-dimensional Lorentz symmetry.  This $Z_2$ symmetry, called Kaluza-Klein
parity, implies the lightest KK particle is stable and can be a good  dark
matter candidate.

6-dimensional UED models are particularly interesting  since they provide
elegant answers to various outstanding questions in the standard model.  For
instance anomaly cancellation in 6D can predict the existence of right-handed
neutrinos and the correct number of fermion generations \cite{Dobrescu:2001ae}.
Proton stability can be attributed to what remains of the 6D Lorentz symmetry
after compactification \cite{Appelquist:2001mj}.

We consider two universal extra dimensions compactified on the chiral square, 
where two adjacent sides are identified
\cite{Dobrescu:2004zi,Burdman:2005sr,Hashimoto:2004xz}.  KK particles are
labeled by two positive integers called `level', $(j, k)$, which represent 
quantization of momentum along the two extra dimensions.  KK particles are odd
under KK parity when $j+k$ is odd while they are even when $j+k$ is even, {\it
i.e.} KK parity is defined as $(-1)^{j+k}$. Their tree-level masses are also
determined by their level $M_{(j,k)} = \sqrt{j^2+k^2}/R$, where $R$ is the radius
of extra dimensions. This model gives rise to  interesting collider
phenomenology, due in part to the presence of new `spinless adjoint' particles,
the uneaten components of extra  dimensional gauge fields.  Furthermore not all
KK masses are integer  multiples of the compactification scale, level-2 KK
particles ($(1,1)$ KK particles), for  instance, have masses a factor $\sqrt{2}$
larger than the level-1 ($1,0$) modes  \footnote{We denote (1,0) state by $\one$
for brevity, especially when it appears as a particle index},  making the
$(1,1)$ KK levels the most easily accessible new particles at the
LHC~\cite{Burdman:2006gy}.

The degeneracy of the modes of one level is lifted by one-loop mass corrections
and by UV-generated boundary terms in a simple implementation of the
6-dimensional standard model on the chiral square, detailed in
\cite{Dobrescu:2004zi,Burdman:2005sr,Hashimoto:2004xz}. 
The 6DSM as an effective theory does not specify the coefficient of the boundary
terms generated by the UV physics.
However, since the one-loop mass
terms from physics below the UV cutoff are enhanced by a logarithmic factor with
respect to the UV boundary terms \cite{Ponton:2005kx,Burdman:2006gy}, 
the latter are neglected throughout this work.
Note that we choose a specific mass spectrum only for concreteness for our
numerical analysis, but our main results do not depend very much on the precise
values of the masses, as long as the hierarchy is not modified.
The one-loop mass corrections result in the spinless adjoints becoming lighter
than their corresponding gauge bosons \cite{Ponton:2005kx}  (different from 5DSM
\cite{Cheng:2002iz} due to existence of spinless adjoints).  By forbidding
decays to pure standard model final states, KK parity forces the (1,0) spinless
adjoints to undergo 3-body decays via heavier $(1,0)$ fermions to pairs of
standard model fermions and another $(1,0)$ mode.  It also ensures the stability
of the hypercharge spinless adjoint, the lightest KK-odd particle (LKP),  making
it a viable dark matter candidate \cite{Dobrescu:2007ec} (see
\cite{Servant:2002aq,Cheng:2002ej} for  comparison with 5-dimensional case). 
Level (1,0) modes pair produced at colliders therefore cascade decay via
spinless  adjoints to the LKP, resulting in events with large numbers of leptons
and missing energy, a very distinctive signal at hadron colliders. Moreover a
non-negligible fraction of multi-lepton events also  contain photons from
one-loop two-body decays of the level (1,0) hypercharge gauge boson
\cite{Dobrescu:2007xf}.  The Feynman rules and mass spectrum are given in
\cite{Burdman:2005sr,Burdman:2006gy,Dobrescu:2007xf}. Following the notation in
these references, we denote the spinless adjoints as $B^\one_H$ for the $U(1)$
hypercharge group and $Z^\one_H$ and $W^{\one\pm}_H$ for  the $SU(2)_W$ weak
group, respectively.

The lowest level (1,0) of KK particles cannot decay into SM particles only, due
to conservation of KK parity, so that they can only be produced in pairs.
Higher level KK states with even KK parity, such as level (2,0) and (1,1)
modes, on the other hand, can couple to SM particles. However, since this kind
of coupling violates KK number, it is only mediated through loop-suppressed
boundary terms (see for example \cite{Burdman:2006gy}). As a result, level
(2,0) or (1,1) states can be produced singly at colliders, but with relatively
small cross sections.

The loop integrals for the mass terms and KK-number violating couplings are
ultraviolet-divergent and need to be regularized by a cut-off. The cut-off scale
$\Lambda$ is estimated by naive dimensional analysis to be $\Lambda \sim
10\times R^{-1}$ \cite{Burdman:2006gy}.

\section{Production and decays of (1,0) particles at the ILC}
\label{sec:prod_decays}

In this section we will consider the production of (1,0) particles in the 6DSM
at the ILC with $\sqrt{s}=1$ TeV as well as their decays. Previously, detailed
studies for level-1 KK-particles have been performed in the context of the 5DSM
\cite{Battaglia:2005zf,Choi:2006mr}. One of the main differences between the
6DSM and the 5DSM are the presence of spinless adjoints. Nevertheless, some of
our results are similar to the 5DSM case due to numerically negligible effects
of spinless adjoints  or their absence in certain processes. However, the
one-loop corrected mass spectrum (see \cite{Cheng:2002iz} and 
\cite{Burdman:2006gy,Ponton:2005kx} for comparison)  is different, and
production of spinless adjoints at the ILC is a new feature of the 6DSM beyond
the 5DSM. Furthermore, previous studies of UED scenarios at linear colliders
have focused on center-of-mass energies of several TeV.  Therefore it is worth
doing a complete analysis in the context of the 6DSM.  For our study we have
implemented the relevant interactions and 1-loop corrected masses of  (1,0)
particles in the 6DSM in {\tt CalcHEP/CompHEP}
\cite{Pukhov:2004ca,Pukhov:1999gg,Boos:2004kh} (see \cite{web} for our model
file). Since we want to compare the signatures with supersymmetry, we are also
making use of the MSSM model file with a neutralino or gravitino as lightest
supersymmetric particle (LSP) for {\tt CalcHEP/CompHEP} \cite{Gorbunov:2001pd}.

Considering the bounds from searches at the Tevatron \cite{Dobrescu:2007xf},
which require $R^{-1} \gtrsim 300$ GeV, it is very unlikely that a 500 GeV
$e^+e^-$ collider could observe any of the KK particles in this model. Therefore
in this work we will focus on a 1~TeV version of the ILC. However, one should
keep in mind that localized boundary terms from UV-completion physics, which are
neglected here, could introduce additional mass splittings between the KK
states. As a result, some KK particles can be lighter than 250 GeV, without
violating the Tevatron limits. We will not explore this possibility further
here.

\subsection{KK fermions}
\label{sec:prod_fermions}
\FIGURE[t]{
\unitlength=1.0 pt
\SetScale{1.0}
\SetWidth{0.7}      
{} 
\allowbreak
\begin{picture}(130,100)(0,0)
\Text(15.0,80.0)[r]{{$e^+$}}
\ArrowLine(40.0,50.0)(20.0,80.0)
\Text(15.0,20.0)[r]{{$e^-$}}
\ArrowLine(20.0,20.0)(40.0,50.0)
\Text(60.0,57.0)[b]{{$\gamma,Z$}}
\Photon(40.0,50.0)(80.0,50.0){3.0}{6}
\Text(105.0,20.0)[l]{{$e^{\one -}$}}
\ArrowLine(80.0,50.0)(100.0,20.0)
\Text(105.0,80.0)[l]{{$e^{\one +}$}}
\ArrowLine(100.0,80.0)(80.0,50.0)
\Text(60.0,0.0)[c]{(a)}
\end{picture} \
\begin{picture}(130,100)(0,0)
\Text(15.0,80.0)[r]{{$e^+$}}
\ArrowLine(60.0,80.0)(20.0,80.0)
\Text(15.0,20.0)[r]{{$e^-$}}
\ArrowLine(20.0,20.0)(60.0,20.0)
\Text(67.0,50.0)[l]{{$B^\one,Z^\one$}}
\Photon(60.0,80.0)(60.0,20.0){3.0}{7}
\Text(105.0,20.0)[l]{{$e^{\one -}$}}
\ArrowLine(60.0,20.0)(100.0,20.0)
\Text(105.0,80.0)[l]{{$e^{\one +}$}}
\ArrowLine(100.0,80.0)(60.0,80.0)
\Text(60.0,0.0)[c]{(b)}
\end{picture} \
\begin{picture}(130,100)(0,0)
\Text(15.0,80.0)[r]{{$e^+$}}
\ArrowLine(60.0,80.0)(20.0,80.0)
\Text(15.0,20.0)[r]{{$e^-$}}
\ArrowLine(20.0,20.0)(60.0,20.0)
\Text(67.0,50.0)[l]{{$B^\one_H,Z^\one_H$}}
\DashLine(60.0,80.0)(60.0,20.0){3}
\Text(105.0,20.0)[l]{{$e^{\one -}$}}
\ArrowLine(60.0,20.0)(100.0,20.0)
\Text(105.0,80.0)[l]{{$e^{\one +}$}}
\ArrowLine(100.0,80.0)(60.0,80.0)
\Text(60.0,0.0)[c]{(c)}
\end{picture} 
\caption{\label{fig:diagrams_ee}
\sl The tree-level Feynman diagrams for KK electron production,
$e^+e^-\rightarrow e^{\one+} {e}^{\one-}$.} }
We first turn to the discussion of (1,0) KK leptons.  The (1,0) muons are
produced via $s$-channel diagrams only (see figure~\ref{fig:diagrams_ee}a) 
\footnote{In principle there are also $s$-channel diagrams with higher KK modes
\cite{Battaglia:2005zf}.  However their couplings to the standard model
particles are 1-loop suppressed and  their contributions are only important when
the center-of-mass energy is close to the threshold. Higher KK bosons in the
6DSM turn out to be almost lepto-phobic \cite{Burdman:2006gy}.}. KK muons
with $\mu^+\mu^- +\met$ signature have been greatly studied in 
\cite{Battaglia:2005zf,Choi:2006mr} in the context of the 5DSM, exhibiting 
a clear difference in the angular distributions of the final state muons in
comparison with supersymmetry.  KK tau-leptons are
similar but can be observed in several modes depending on different $\tau$ decay
channels.
However, unlike the 5DSM where both $SU(2)_W$ singlet and doublet KK leptons
always decay into  a standard model lepton and the LKP $B_\mu^1$, in the 6DSM a
doublet (1,0) KK lepton can decay into one of several lighter (1,0) particles: 
47\% of the time for $W^{\one\pm}_H$, 23.5\% for $Z^\one_H$  (which is
equivalent to $W_H^{\one 3}$ due to the small Weinberg angle for KK states),  20\%
of the time for $B^\one_H$,  and only 9.3\% for $B^\one_\mu$
\cite{Dobrescu:2007xf}.  Using formulas for the decay withs of KK-leptons given
in \cite{Dobrescu:2007xf}, we found that the singlet (1,0) lepton decays into
$B^\one_H$ 81\% of the time and into $B^\one_\mu$ with 19\% branching fraction. 
The existence of spinless
adjoints in this model greatly diversifies  the decay modes of KK leptons,
leading to richer phenomenology. For instance, a (1,0) KK-lepton can decay into
$B^\one_\mu$ plus a lepton, with the $B^\one_\mu$ continuing to decay into
$B^\one_H$ emitting two leptons or a photon.
\FIGURE[t]{
\centerline{\epsfig{file=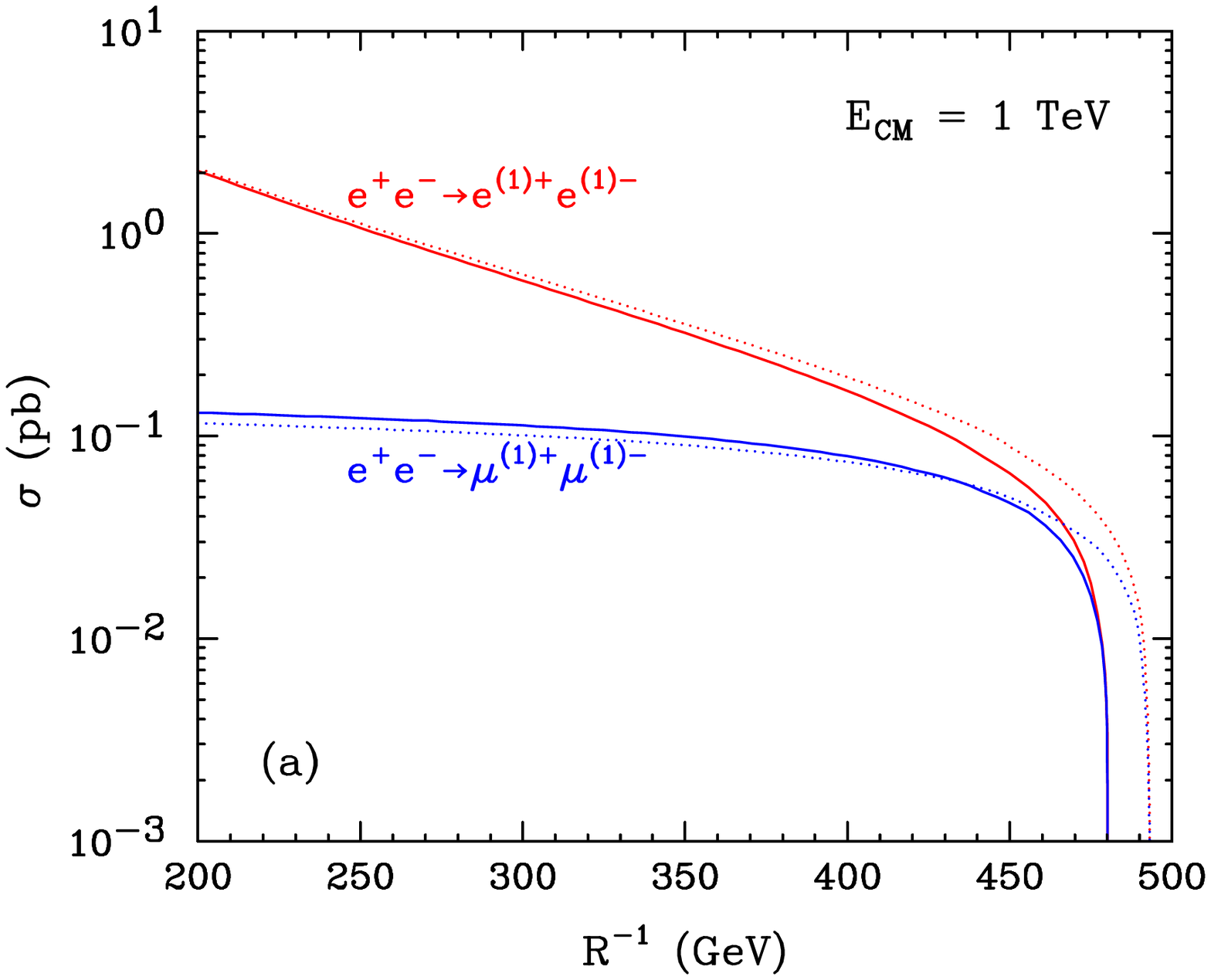,width=7.4cm} \hspace{0.2cm}
\epsfig{file=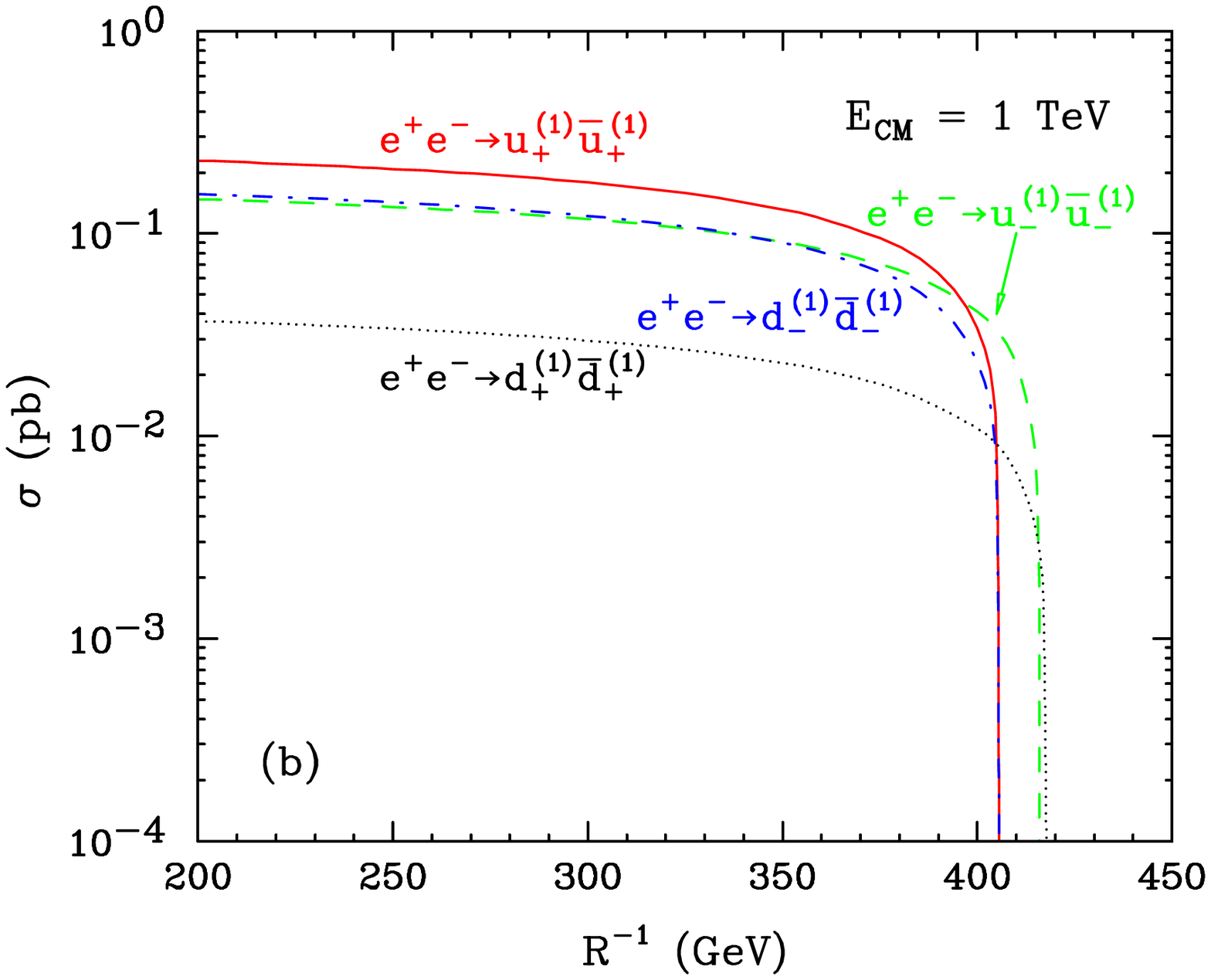,width=7.4cm} 
\caption{\sl ISR-corrected production cross sections of 
(a) (1,0) KK leptons ($e^\one$ in red $\mu^\one$ in blue) and 
(b) (1,0) KK quarks, as a function of $R^{-1}$.
Solid (dotted) lines correspond to $SU(2)_W$-doublets (singlets).}
\label{fig:fermions}}}

The case of (1,0) electrons is more interesting. Their production can also
proceed  through the $t$-channel diagrams as shown in
figures~\ref{fig:diagrams_ee}b and~\ref{fig:diagrams_ee}c,  leading to larger
cross sections. The pair production cross section for (1,0) leptons are shown in
figure~\ref{fig:fermions}. We found that diagrams with t-channel spinless
adjoints contribute little to the total cross section. Angular distributions and
spin correlations can provide important information about the spins of the
particles in the $t$-channel and in the decays, and thus allow to discriminate
this model from 5DSM or supersymmetry. However, due to the kinematics of the
dominant $t$-channel exchange, the final state level-0 electron preferably goes
in the forward direction, thus partially obscuring the spin correlation effects.
On the other hand, utilizing polarized  $e^+$ or $e^-$ beams allows us to obtain
additional information about the spin structure (see Ref.~\cite{power} for a
review on the role of polarized beams at ILC), as will be discussed in more
detail in the next section.

KK quarks are produced in the same way as KK muons through photon and $Z$
exchange in the $s$-channel. In the 5DSM, doublet and singlet KK quarks have
different decay patterns:  singlet KK quarks always decay into the $B^1_\mu$ LKP
while the doublet KK quarks decays into  $W^{1 \pm}_\mu$ or $Z_\mu^1$. In the
6DSM, the dominant decay modes of both singlet and doublet KK quarks are into
$G_H^\one$ (see Table 3 in \cite{Dobrescu:2007xf} for branching fractions). The
production cross sections of KK quarks plotted in figure~\ref{fig:fermions}b are
of the same order as for KK-muons, but show a somewhat lower mass reach than for
the case of KK leptons, since the KK-quarks receive large positive mass
corrections. KK quark production is the only way to study the properties of the
$G_H^\one$ spin-0 color-octet at the ILC, since the decay of KK quarks is the
only way to produce it at lepton colliders.

\subsection{KK Bosons}
\label{sec:prod_bosons}

The ISR-corrected production cross sections for (1,0) electroweak gauge bosons 
are shown in figure \ref{fig:bosons}a.  The production mechanism is identical to
that in the standard model, replacing the standard model  fermion with the
corresponding KK mode, for example, as shown in figure \ref{fig:diagrams_WW}
for  $W_\mu^{\one\pm}$ production  (there are no $s$-channel diagrams for the
production of neutral bosons).  The branching fractions of the weak gauge
bosons  ($W_\mu^{\one\pm}$ and $Z_\mu^\one$ ($W_\mu^{\one 3}$)) are the same as
those in the 5DSM (100\% to $SU(2)_W$ doublet leptons).  
However, while in the 5DSM the KK hypercharge gauge boson ($B^1_\mu$) is stable,
in the 6DSM the $B^\one_\mu$ decays into the $B_H^\one$ and SM particles. It
decays into a photon plus $B_H^\one$ 34\% of the time,  and into two leptons
plus $B_H^\one$ with 21.3\% branching fraction per generation.  This resembles
the phenomenology of the lightest neutralino with  a gravitino LSP. The pair
production cross section for $B^\one_\mu$ is relatively large,  $\sim$ 100 fb
for $R^{-1} \lesssim 500$ GeV. We will discuss the $2\gamma+ \met$ signature in
the next section in detail.

\FIGURE[t]{
\unitlength=1.0 pt
\SetScale{1.0}
\SetWidth{0.7}      
{} 
\allowbreak
\begin{picture}(120,100)(0,0)
\Text(15.0,80.0)[r]{{$e^+$}}
\ArrowLine(40.0,50.0)(20.0,80.0)
\Text(15.0,20.0)[r]{{$e^-$}}
\ArrowLine(20.0,20.0)(40.0,50.0)
\Text(60.0,57.0)[b]{{$\gamma,Z$}}
\Photon(40.0,50.0)(80.0,50.0){3.0}{6}
\Text(105.0,20.0)[l]{{$W^{\one-}_\mu$}}
\Photon(80.0,50.0)(100.0,20.0){3.0}{5}
\Text(105.0,80.0)[l]{{$W^{\one-}_\mu$}}
\Photon(80.0,50.0)(100.0,80.0){3.0}{5}
\Text(60.0,0.0)[c]{(a)}
\end{picture} 
\hspace*{1cm}
\begin{picture}(120,100)(0,0)
\Text(15.0,80.0)[r]{{$e^+$}}
\ArrowLine(60.0,80.0)(20.0,80.0)
\Text(15.0,20.0)[r]{{$e^-$}}
\ArrowLine(20.0,20.0)(60.0,20.0)
\Text(67.0,50.0)[l]{{$\nu_e^{\one}$}}
\ArrowLine(60.0,20.0)(60.0,80.0)
\Text(105.0,20.0)[l]{{$W^{\one-}_\mu$}}
\Photon(60.0,20.0)(100.0,20.0){3.0}{6}
\Text(105.0,80.0)[l]{{$W^{\one+}_\mu$}}
\Photon(60.0,80.0)(100.0,80.0){3.0}{6}
\Text(60.0,0.0)[c]{(b)}
\end{picture} \
\caption{\label{fig:diagrams_WW}
\sl The tree-level Feynman diagrams for $W^{\one-}_\mu W^{\one+}_\mu$ production}}
Another important distinction is the production of spinless adjoints, for which  we
show our results in figure \ref{fig:bosons}b. Production of
$W_H^{\one+}W_H^{\one-}$ is mediated by photon and $Z$ exchange in the
$s$-channel and  KK-neutrino exchange in the $t$-channel, as in the case of
$W_\mu^{\one+}W_\mu^{\one-}$, while only $t$- or $u$-channel diagrams contribute
for neutral bosons.  The spinless adjoints have rather small cross sections
except for charged pair production, due to the $s$-channel contribution. As a
result of the approximate degeneracy in the mass spectrum, two-body decays are not
allowed and  the main decay mode of weak spinless adjoints is into lepton pairs
plus $B_H^\one$. 
\FIGURE[t]{ 
\centerline{\epsfig{file=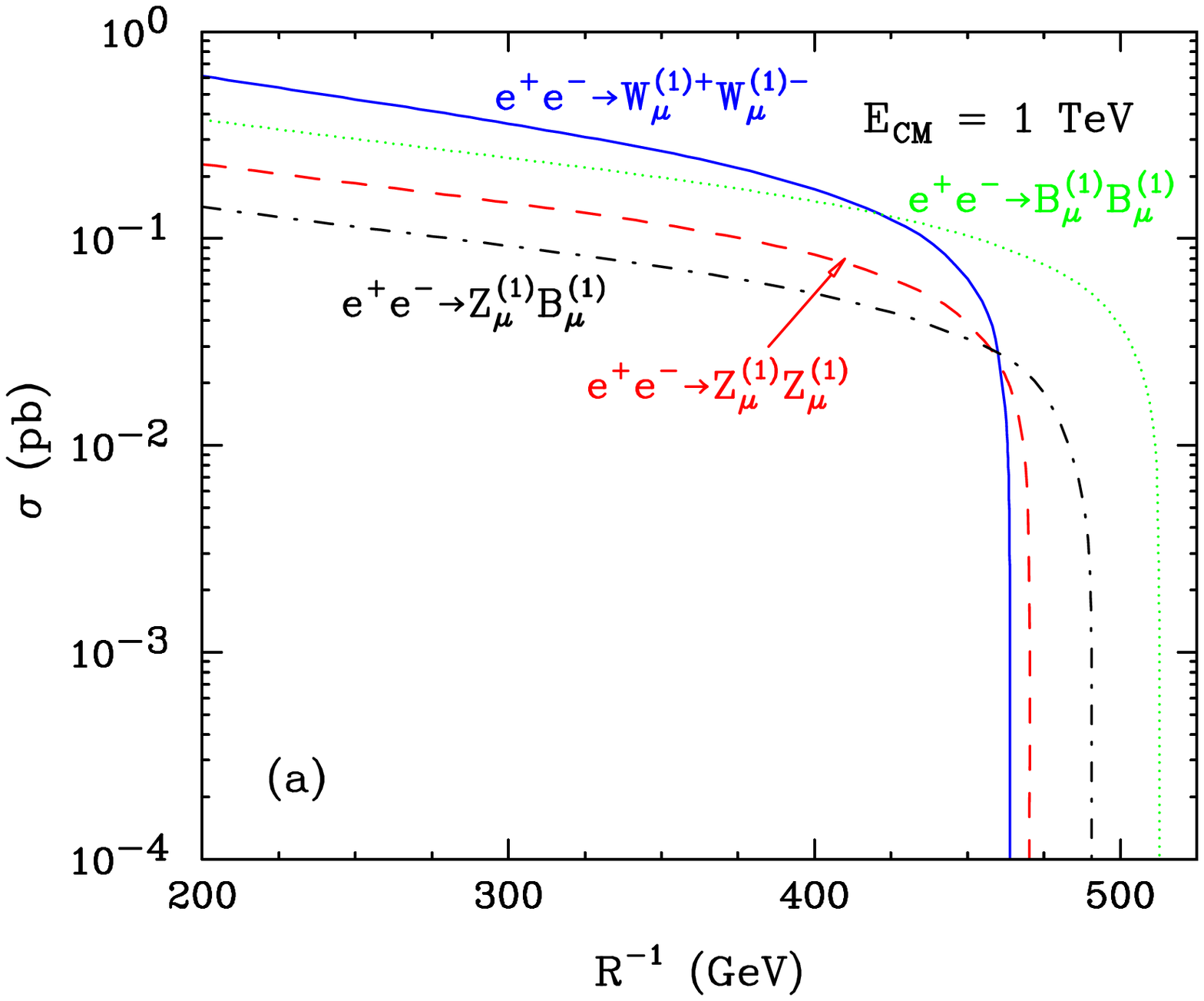,width=7.4cm} \hspace{0.2cm}
\epsfig{file=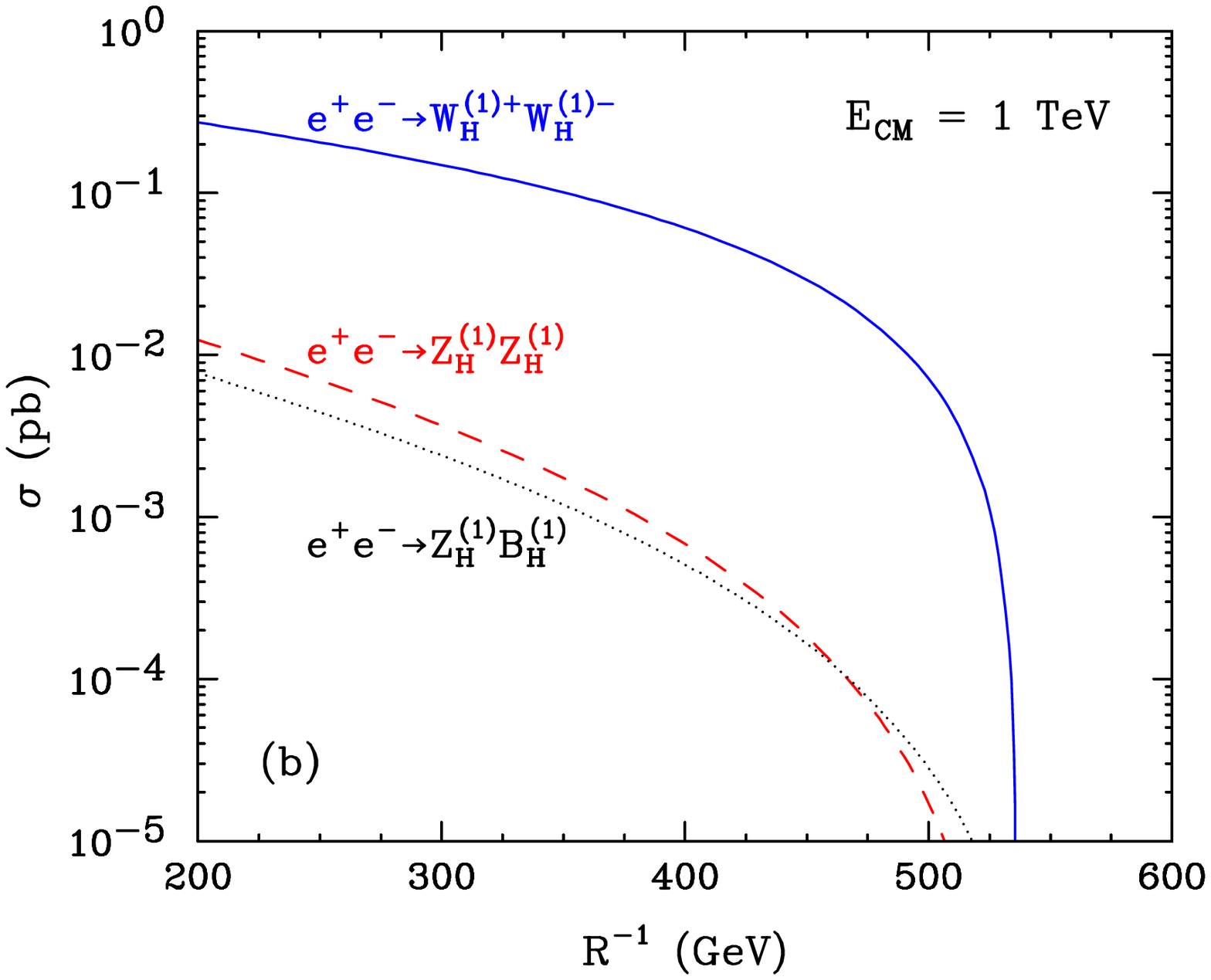,width=7.4cm}
\caption{\sl ISR-corrected production
cross sections of  (a) (1,0) KK vector bosons and  (b) (1,0) spinless adjoints,
as a function of $R^{-1}$.} \label{fig:bosons}}}

In figure~\ref{fig:transition} we summarize the decay patterns of (1,0) 
particles of the 6DSM in a pictorial way in comparison with the 5DSM
\cite{Cheng:2002ab}. There are two separate groups of particles: one (left in
red) arising in both 5DSM and 6DSM, and the other (right in blue) that exists
only in the 6DSM.  These additional states are all spinless adjoints that are
lighter than the $B^\one_\mu$. One important consequence of this is  that
(1,0) fermions (circled) decay into these spinless adjoints with  non-negligible
branching fractions, thus completely changing the collider phenomenology. 
\FIGURE[t]{
\epsfig{file=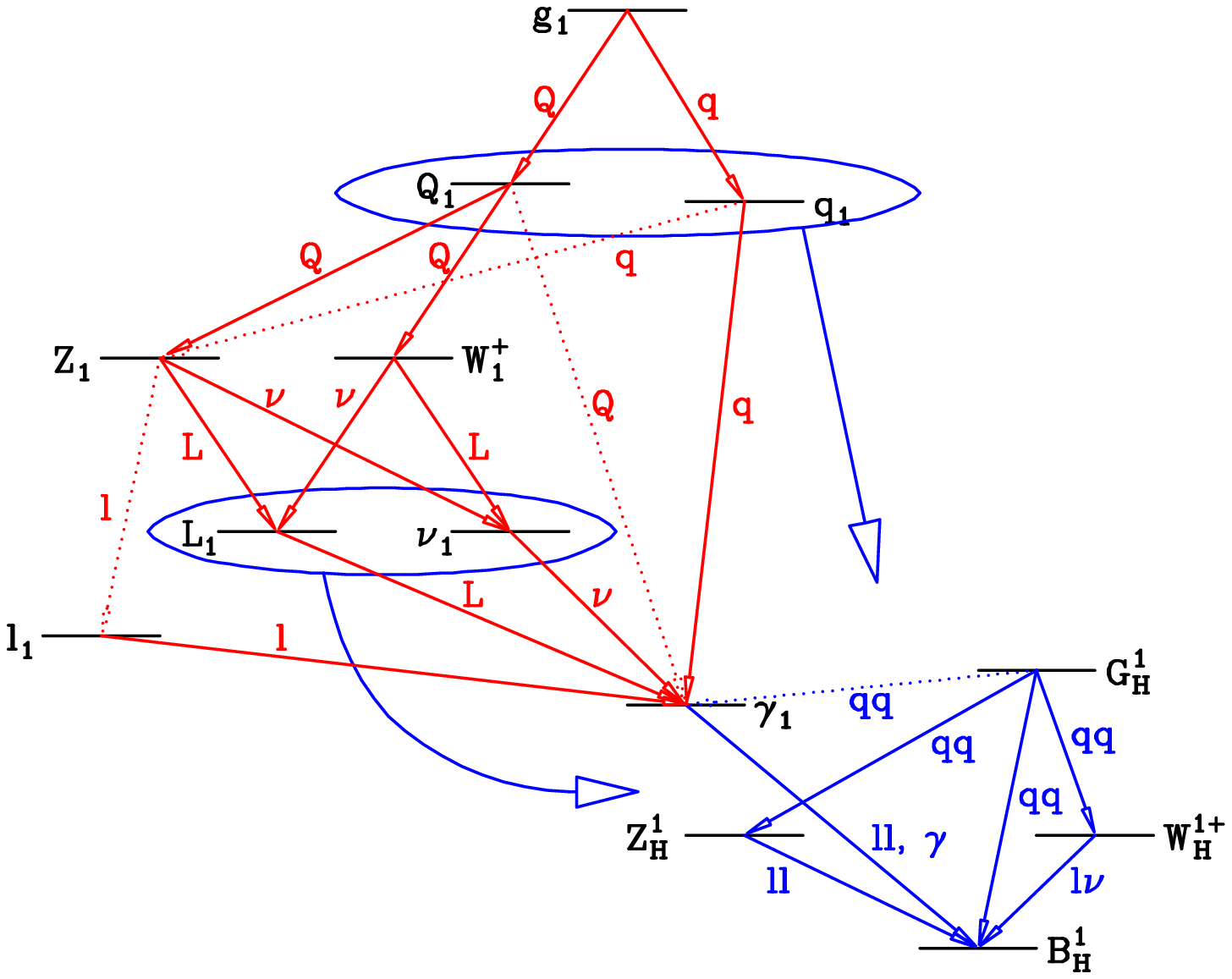, width=10cm}
\caption{\sl Schematic diagram for the decays of (1,0) KK particles. The typical
particle spectrum and decay patterns of the 5DSM are shown in red, while the
6DSM encompasses the particles and decay modes depicted both in red and blue.}
\label{fig:transition}}
%
%

\section{Determination of particle properties and discrimination between 5DSM and 6DSM}
\label{sec:signatures}

In this section we consider specific signatures involving leptons and photons that are 
produced in the decay of electroweak (1,0) particles.

\subsection{Spin determination in KK-Lepton production and decays}
\label{sec:leptons}

In 5DSM, the lightest KK-odd particle (LKP) is typically a vector boson
($B^\one_\mu$) while in 6DSM it is a scalar ($B^\one_H$). For testing and
discriminating these models it is therefore essential to determine the spin of
the LKP\footnote{In general, the production cross sections are different between
the models and may be used as a discriminator. However, the cross sections and
branching fractions depend sensitively on the spectrum of the complete model.
In this study we instead focus on spin determination from shape information of
the final state distributions, which does not require reconstruction of the
whole mass spectrum of the model.}. This can be done by studying the decays of KK-electrons, as will be
discussed in this section. The KK-excitations of right-handed electrons,
$e^\one_-$, decay directly into $B^\one_\mu$ or $B^\one_H$ and an electron.
Therefore the pair production of $e^{\one+}_- e^{\one-}_-$ gives a very simple
signature of $e^+e^-$ and missing energy. By looking at the angular correlations
of the final state $e^+/e^-$, one can learn something about the spin of the
unreconstructed LKP. The use of polarized beams plays an important role here, as
will be shown in the following.

For instance, one can study pair production of KK-electrons, $e^{\one+}_-
e^{\one-}_-$, near their kinematical threshold. In this case, the KK-electrons
are produced almost at rest. Owing to the $t$-channel diagrams for the pair
production process, the production cross section is enhanced for a right-handed
polarized electron beam, and a left-handed polarized positron beam. In this
case, the initial $e^+e^-$ state has a total spin of 1, so that the two produced
KK-electrons need to have a combined total spin pointing in the same direction
(the direction of the incoming $e^-$). Due to the helicity structure of the
KK-electron couplings, the decay $e^{\one\pm}_- \to e^\pm \, B^\one_\mu$
produces left-handed electrons (right-handed positrons) for transverse
polarization of the $B^\one_\mu$, whereas $e^{\one\pm}_- \to e^\pm \, B^\one_H$
leads to right-handed electrons (left-handed positrons). As a consequence,
angular momentum conservation mandates that for the decay into $B^\one_\mu$,
the outgoing $e^-$ goes preferentially in the backward direction with respect to
the incoming $e^-$ beam, whereas for the decay into the scalar $B^\one_H$, the 
$e^-$ dominantly goes in the forward direction.

In the following, we will be studying three cases:
\paragraph{6DSM:}
a scenario of a model with two universal extra dimensions and $R^{-1} = 300$
GeV. The masses and branching ratios of the  KK-electrons and hypercharge bosons
in this scenario are
\begin{equation}
\begin{aligned}
M_{B^\one_\mu} &= 292.0 \gev, \\
M_{B^\one_H} &= 256.5 \gev, \\
M_{e^\one_-} &= 304.5 \gev, &
	&{\rm BR}[e^{\one-}_- \to e^- B^\one_\mu] = 19\%, \\[-1ex] 
\Gamma_{e^\one_-} &= 0.040 \gev, &
	&{\rm BR}[e^{\one-}_- \to e^- B^\one_H] = 81\%, \\
M_{e^\one_+} &= 312.0 \gev, &
	&{\rm BR}[e^{\one-}_+ \to e^- B^\one_\mu] = 12\%, \\[-1ex] 
\Gamma_{e^\one_+} &= 0.038 \gev, &
	&{\rm BR}[e^{\one-}_+ \to e^- B^\one_H] = 27\%, \\[-1ex] 
&&	&{\rm BR}[e^{\one-}_+ \to e^- W^{\one0}_H, \nu_e W^{\one-}_H] = 60\%.
\end{aligned}
\end{equation}
Since the $B^\one_\mu$ is heavier than the $B^\one_H$, it will decay into the
latter plus a photon or a pair of leptons. By demanding a final state with a
electron-positron pair and missing energy only, one can therefore select a
sample of events  which practically only contains direct decays to the LKP
$B^\one_H$.

\paragraph{5DSM:}
a scenario of a model with one universal extra dimensions with KK-electron and
KK-$B$-boson masses chosen to match the kinematics of the 6DSM scenario:
\begin{equation}
\begin{aligned}
M_{B^\one_\mu} &= 256.5 \gev, \\
M_{e^\one_-} &= 304.5 \gev, &
	&{\rm BR}[e^{\one-}_- \to e^- B^\one_\mu] = 100\%, \\[-1ex] 
\Gamma_{e^\one_-} &= 0.11 \gev, 
\\
M_{e^\one_+} &= 312.0 \gev, &
	&{\rm BR}[e^{\one-}_+ \to e^- B^\one_\mu] = 33\%, \\[-1ex] 
\Gamma_{e^\one_+} &= 0.11 \gev, &
	&{\rm BR}[e^{\one-}_+ \to e^- W^{\one0}_\mu, \nu_e W^{\one-}_\mu] =
	67\%.
\end{aligned}
\end{equation}

\paragraph{SUSY:}
for completeness, we also consider a MSSM scenario, where the masses of the
selectrons and neutralinos have been chosen to be equal to the masses of the
KK-electrons and -bosons in the 6DSM scenario.\\[1ex]
\begin{equation}
\begin{aligned}
M_{\tilde{\chi}^0_2} &= 292.0 \gev, \\
M_{\tilde{\chi}^0_1} &= 256.5 \gev, \\
M_{\tilde{e}_R} &= 304.5 \gev, &
	&{\rm BR}[\tilde{e}_R \to e^- \tilde{\chi}^0_1] = 100\%, \\[-1ex]  
\Gamma_{\tilde{e}_R} &= 0.13 \gev.
\end{aligned}
\end{equation}
The contribution of the L-selectron has been disregarded in the SUSY scenario,
since it can be significantly heavier than the R-selectron.

\paragraph{Idealized picture: no background, 100\% polarization.}

We have generated signal cross sections and distributions for the three cases
with {\tt CompHEP 4.4} and {\tt CalcHEP 2.5}. In a first step, only the
signal process without backgrounds and with 100\% polarization is considered.

As can be seen in figure~\ref{fg:thr}a, near the production threshold of $2
\times 304.5$ GeV $= 609$ GeV, the 6DSM process with scalar adjoints in the
final state is dominantly emitting the $e^-$ in the forward direction, while the
differential cross section vanishes in the backward  direction.
\begin{center}
\FIGURE[t]{
\centerline{\epsfig{figure=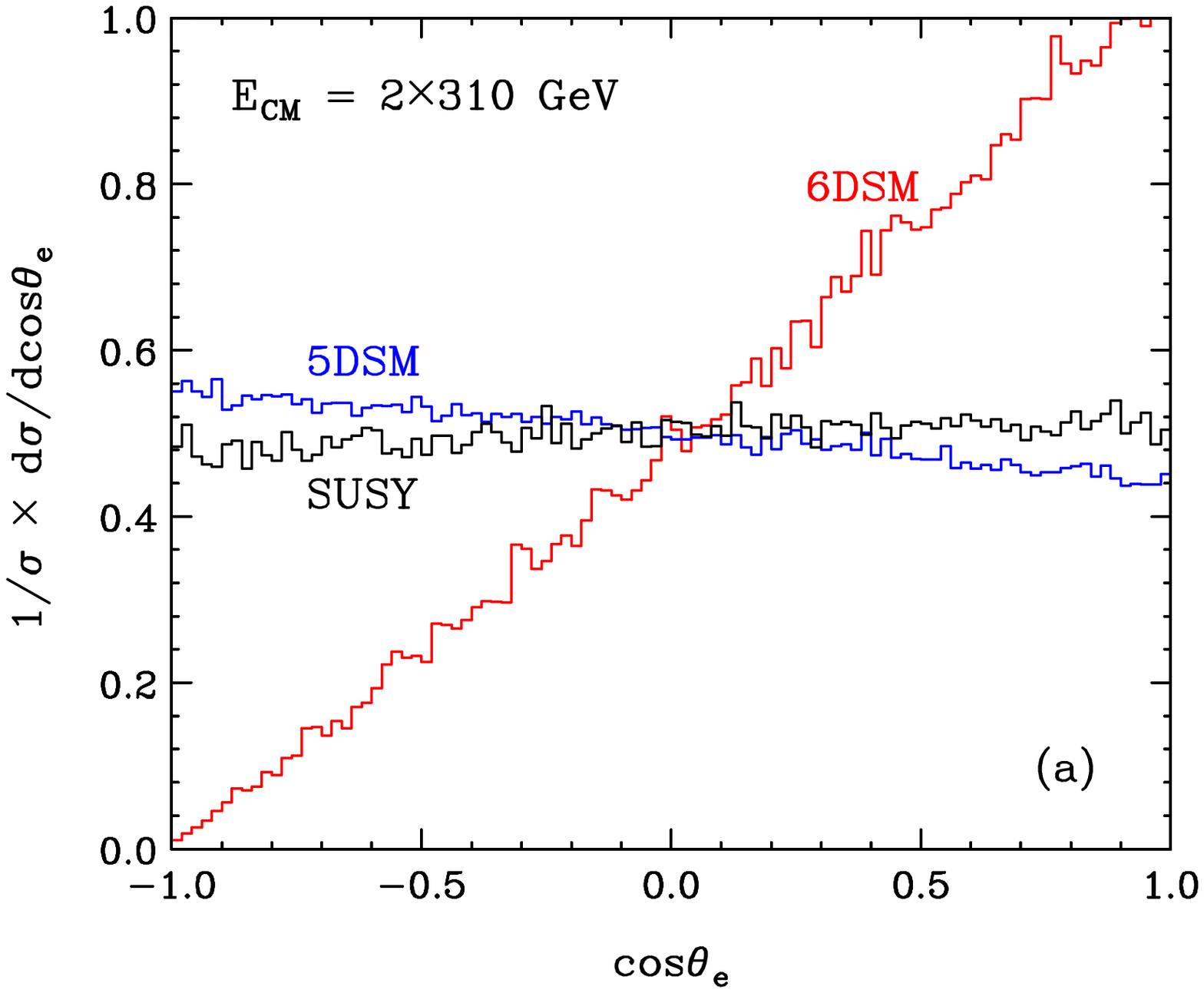, width=7.4cm} \hspace{0.2cm}
\epsfig{figure=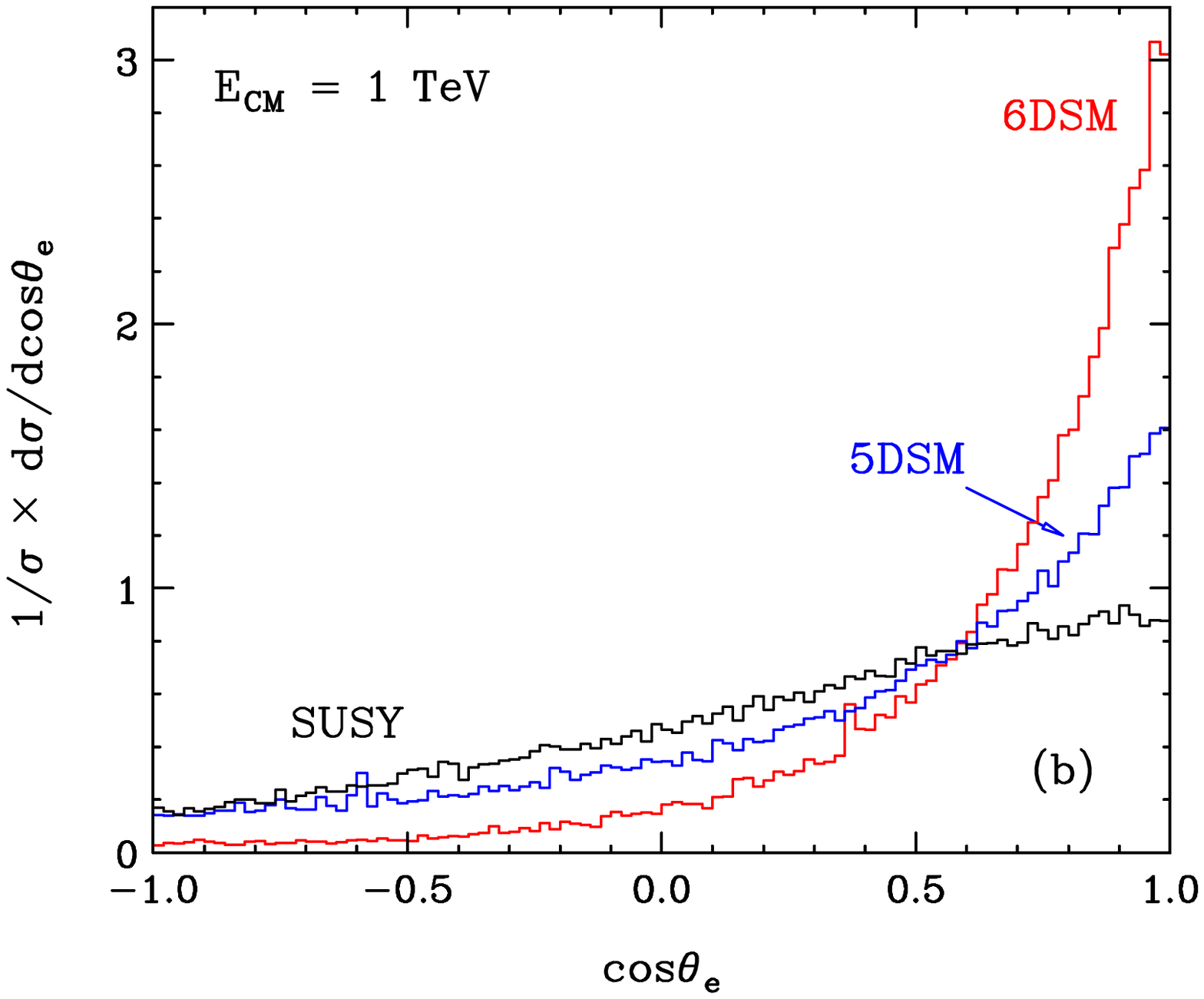, width=7.4cm,height=6.05cm} 
\caption{\sl Angular dependence of the decay electron originating from pair
production of heavy right-handed electron partners 
for 100\% polarized $e_{\rm R}^-$ and $e_{\rm L}^+$ beams 
(a) near their production threshold and (b) at higher center-of-mass energy
$\sqrt{s} = 1$ TeV $\gg 2 \times m_{e \rm partner}$. 
These plots compare three
extensions of the standard model, where the heavy electron partners decays into
an electron plus a stable weakly interacting particle. This stable particle can 
be a scalar (6DSM: $e^+e^- \to e_+^{\one+} e_-^{\one -} \to e^+e^- B_H^\one B_H^\one$), 
vector (5DSM: $e^+e^- \to e_+^{1+} e_-^{1 -} \to e^+e^- B_\mu^1 B_\mu^1$) or 
Majorana fermion (SUSY: $e^+e^- \to \tilde{e}_R \tilde{e}_R^\ast \to e^+e^- \tilde{\chi}_1^0\tilde{\chi}_1^0$). 
The distributions are normalized to unity. 
}
\label{fg:thr}
}}
\end{center}
On the other hand, in the 5DSM case where the LKP is a vector, the final state
electron prefers to go in the backward direction. Due to the contribution of
longitudinally polarized $B^\one_\mu$ bosons, this is however a relatively small
effect. Finally, for the case of supersymmetry (SUSY), the angular distribution
is exactly flat owing to scalar nature of the selectrons, see figure~\ref{fg:thr}a.

Since a future $e^+e^-$ collider will spend most time running at specific design
center-of-mass energies, it is interesting to see how much of these spin
dependence effects survive at higher values of $\sqrt{s}$, {\it i.e.} not close
to the threshold. This is shown in figure~\ref{fg:thr}b. Due to the sizable
boost of the heavy electron partners (KK-electrons or selectrons) the angular
distributions for all models get deformed toward the forward region.
Nevertheless, a sizable difference between the three cases remains.
More information can be gained by also looking at the energy distribution of the
final state $e^+/e^-$. Again as a result of the scalar nature of the selectron,
the final state electron energy distribution in the SUSY case is perfectly flat,
with a minimum and maximum given by the selectron and neutralino masses
\begin{equation}
E_{\rm min,max} = \frac{\sqrt{s}}{4} \;
\frac{m_{\tilde{e}}^2-m_{\tilde{\chi}^0_1}^2}{m_{\tilde{e}}^2}
\; \left( 1 \pm \sqrt{1- \frac{4 m_{\tilde{e}}^2}{s}} \right). \label{eq:edges}
\end{equation}
Since these lower and upper endpoints are defined solely through kinematics,
they are identical for all three models (5DSM, 6DSM and SUSY), provided that the
partners of the electrons and gauge bosons have the same masses in all models.

However, the forward peak due to angular correlations in the 6DSM model leads to
a peak at the upper end of the electron energy distribution, see
figure~\ref{fg:edist}. The reason is that the forward direction of the electron
corresponds to the upper end of its energy distribution, since in this
kinematical configuration the boost of the KK-electron and the momentum of
the electron are collinear. On the other hand, the electron energy distribution
for the 5DSM model exhibits a slight slope toward the lower end.
\FIGURE[t]{
\centerline{\epsfig{figure=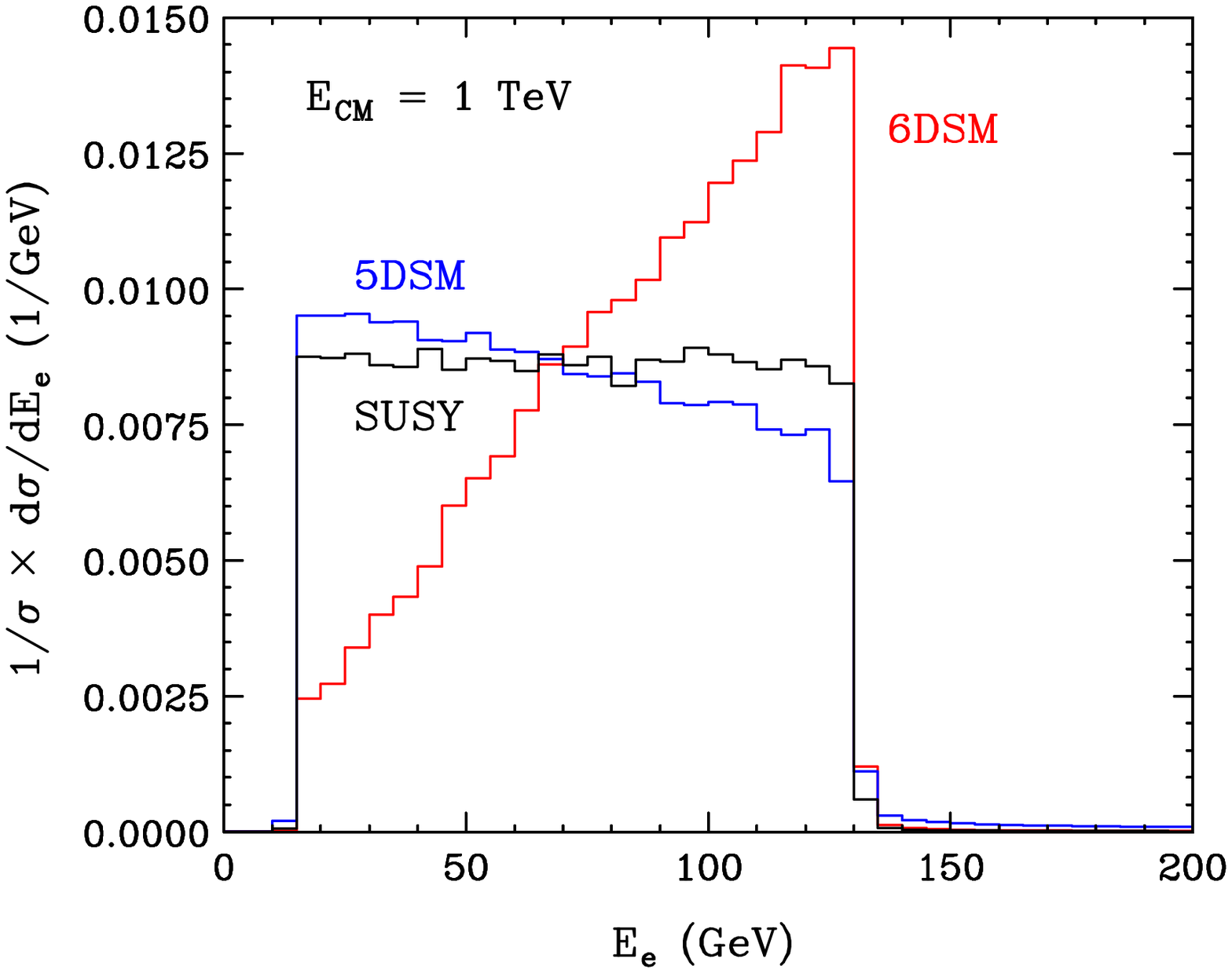, width=7.4cm} 
\caption[]{\sl Energy dependence of the decay electron originating from pair
production of heavy right-handed electron partners, for 100\% polarized $e_{\rm
R}^-$ and $e_{\rm L}^+$ beams. The notation is as in figure~\ref{fg:thr}.}
\label{fg:edist}
}}

\paragraph{Realistic picture: standard model backgrounds and
realistic beam polarization.}

In the following,
we assume 80\% polarization of the elecron beam and 50\% polarization of the
positron beam. As pointed out above, right-handed $e^-$ polarization and
left-handed $e^+$ polarization enhance the signal of $e^{\one+}_-
e^{\one-}_-$ production. In addition, this polarization combination also
suppresses standard model background from $W^+W^-$ production.

Without cuts, standard model backgrounds can be large and overwhelm the signal.
The main standard model backgrounds for the signature of $e^+e^- + \Eslash$
(where $\Eslash$ stands for missing energy) come from the processes  $e^+e^- \to
e^+e^- \nu_e \nu_e$ (which includes $W^+W^-$ and $ZZ$ production) and $e^+e^-
\to e^+e^- \gamma \gamma \to e^+e^- e^+e^-$. In the second process, the incoming
$e^+/e^-$ each radiate a near-collinear photon and proceed down the beam pipe.
The two photons collide and generate an $e^+e^-$ pair that is seen in the
detector, while a missing energy signature is created by a small but
non-negligible $p_\perp$ kick of the $e^\pm$ that are lost in the beam pipe.

In the following these standard model backgrounds are studied and it is shown
how they can be reduced with simple selection cuts. To parametrize the detector
acceptance, we demand the detected $e^+$ and $e^-$ have a polar angle
$\theta_{e^\pm}$ in the central detector region, $|\cos \theta_{e^\pm}| < 0.9$, 
a minimum energy $E_{e^\pm} > 5$ GeV and  pair invariant mass $m_{ee} > 10$ GeV.
For the two-photon background we assume that the blind region around the beam
pipe extends to $5^{\circ}$ (87 mrad).

With this setup one finds the cross sections in the second line of the following
table. Note that the standard model backgrounds are ill-defined without the
minimal detector acceptance cuts, since they would diverge due to $t$-channel
singularities.

Even without any selection cuts, the expected UED signal is already of the same
order as the standard model backgrounds. In the SUSY case, however, the
cross section is smaller, and thus we will apply additional cuts that are
sufficient to ensure a clean SUSY signal.

Due to the heavy particles involved in the signal processes, they lead to larger
values for the missing energy and the $e^+e^-$ invariant mass than the standard
model backgrounds. Therefore we apply the cuts $\Eslash > 0.7 \sqrt{s}$ and
$m_{ee} > 100$ GeV. Furthermore, a lower limit on the total transverse momentum,
$p_\perp > 10$ GeV (where the electron momenta have been summed vectorially), is
effective against the two-photon background. After cuts one obtains the cross
sections in the last line of the table.

\vspace{1em}
\noindent
\begin{tabular}{l|rrr}
\hline
 & Signal $e^+e^- B^\one_H B^\one_H$ & $e^+e^- B^\one_\mu B^\one_\nu \quad$ & 
 	$e^+e^- \tilde{\chi}^0_1 \tilde{\chi}^0_1 \qquad$  \\
\hline
No cuts & 1.18(3)\phantom{00} pb \phantom{(99\%)} 
	& 2.42(2)\phantom{0} pb \phantom{(99\%)} 
	& 116(2) \phantom{.0}fb \phantom{(99\%)} 
\\
Acceptance cuts & 0.646(15) pb (55\%) 
	& 1.61(2)\phantom{0} pb (66\%) 
	& 98(2) \phantom{.0}fb (85\%) \\
Selection cuts & 0.518(2)\phantom{0} pb (44\%)
	& 0.703(3) pb (29\%) 
	& 53.9(4) fb (46\%) \\
\hline
\end{tabular}\\
\begin{tabular}{l|rr}
\hline
 & Background $e^+e^- \nu_e \nu_e$ & 
 	$\gamma\gamma  \to e^+e^-$ \\
\hline
Acceptance cuts & 
 	9.07(8)\phantom{0} fb & 1.5(2) pb \\
Selection cuts & 
	0.263(2) fb & 3.9(2) fb \\
\hline
\end{tabular}
\vspace{1em}

The angular and energy distributions after cuts are shown in
figure~\ref{fg:angdist_cut}. The signal cross sections have been normalized to
unity, while the standard model background has been scaled in the same way as
the 6DSM signal cross section, {\it i.e.} the figure reflects the
signal-to-background ratio that is expected for the 6DSM.
It can be seen that for the large expected signal
cross sections in the UED scenarios the standard model background  is totally
negligible. While the shape of the angular distribution is not affected by the
selection cuts, the energy distribution is distorted near its lower end.
Nevertheless, a clear distinction between the three models is possible from the
upper end of the energy distribution.

\FIGURE[t]{
\centerline{
\epsfig{figure=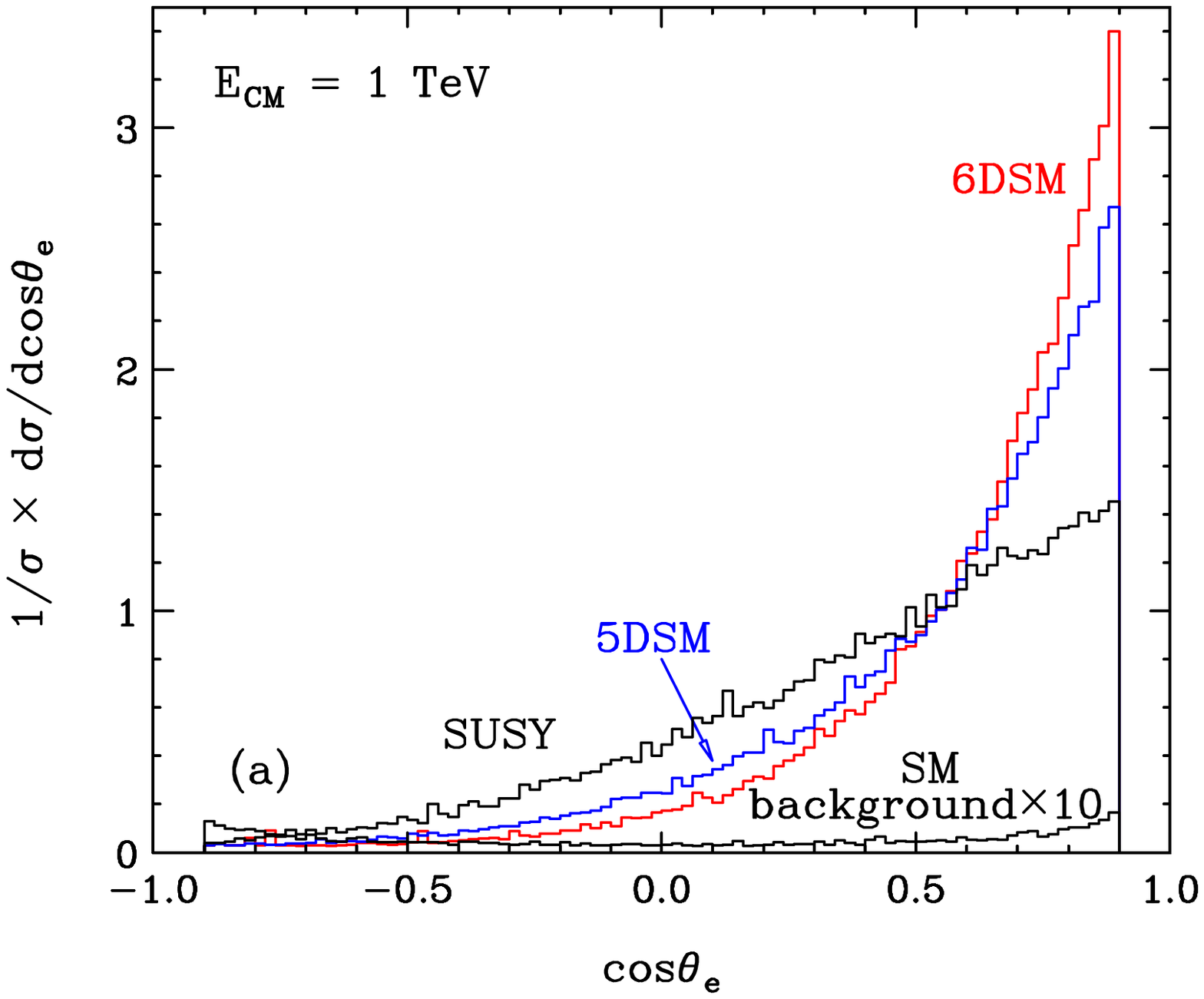, width=7.4cm} \hspace{0.2cm}
\epsfig{figure=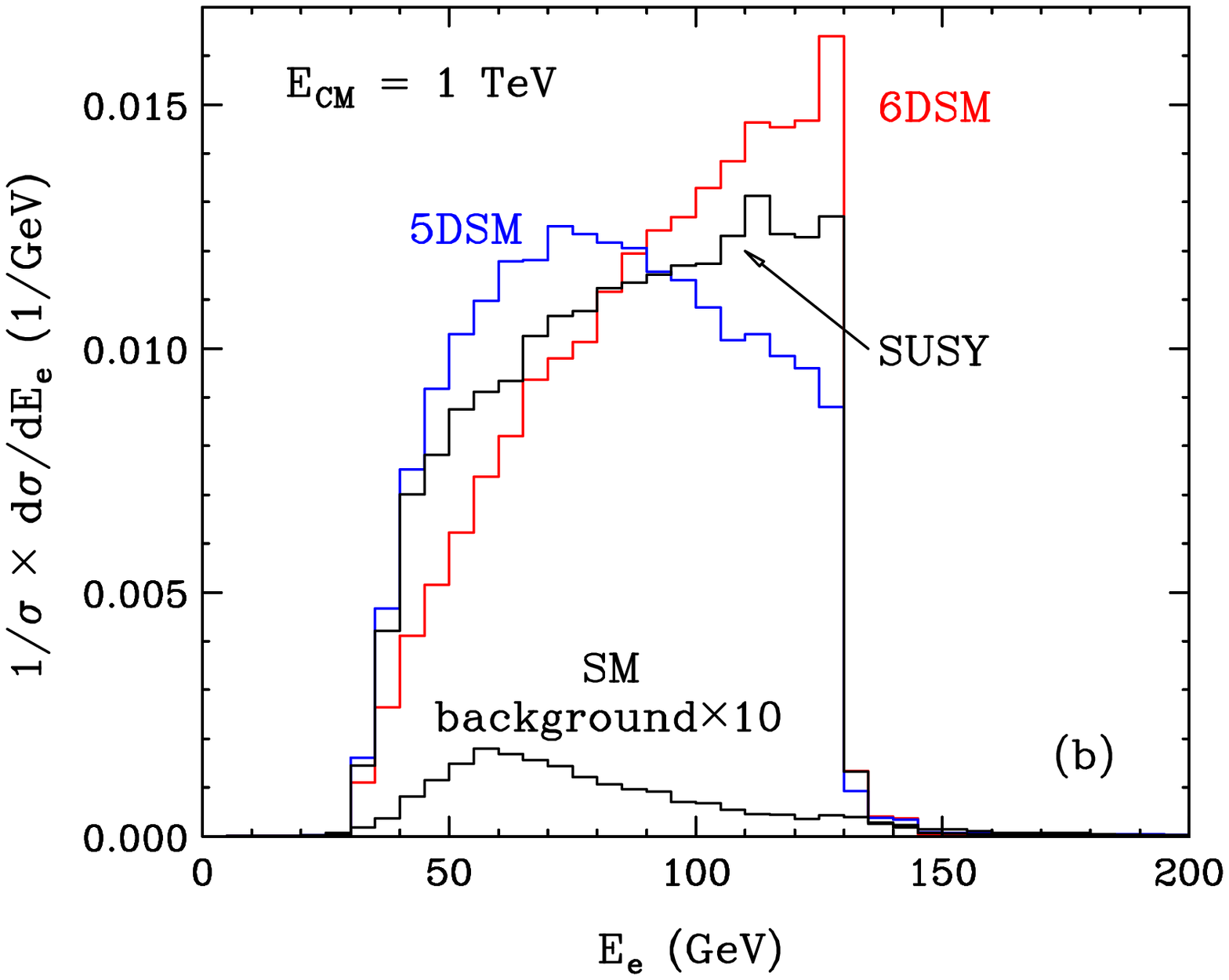, width=7.4cm, height=6.15cm} 
\caption[]{\sl (a) Angular distribution and (b) energy distribution 
of the decay electron originating from pair
production of heavy right-handed electron partners at $\sqrt{s} = 1$ TeV, after
application of cuts to reduce standard model background.
The distributions have been normalized to unity for the three new physics
models. Also shown is the standard model background for the 6DSM case, scaled by
a factor 10 to make it visible.}
\label{fg:angdist_cut}
}}

Assuming 500~fb$^{-1}$ luminosity at $\sqrt{s} = 1$ TeV, and a signal
cross section of 518 fb (which corresponds to the 6DSM case) for all three
models (6DSM, 5DSM and SUSY), the 5DSM and 6DSM spin structure can be
distinguished by about 12 standard deviations from the angular distribution
shape. Here only statistical errors have been taken into account, using a binned
$\chi^2$ test.
The SUSY spin assignments can be discriminated from the 6DSM case by about 37
standard deviations. When instead considering the energy distributions, 5DSM and
6DSM can be separated by about 32 standard deviations, while SUSY and 6DSM can
be distinguished by 18 standard deviations.

In other words, for a separation of the three scenarios at 95\% confidence
level, 2 fb$^{-1}$ luminosity at $\sqrt{s} = 1$ TeV is sufficient. Even if the
cross section after cuts is only 54 fb, as in the SUSY case, 19 fb$^{-1}$
luminosity suffices for the discrimination.

\subsection{Photonic decays of hypercharge bosons}
\label{sec:photons}

As discussed in Section \ref{sec:prod_bosons}, $B_\mu^\one$ pairs are produced
through  $e^\one_+$ and $e^\one_-$ exchange in the $t$-channel while
$\tilde{\chi}^1_0$ pair production is mediated by $s$-channel  $Z$ boson as well. In the
6DSM, they can decay into a photon plus the invisible LKP, $B_H^\one$, with a
sizable branching fraction of 34\% \cite{Dobrescu:2007xf}. Thus they lead to
the characteristic final state signature of two photons plus missing energy,
$2\gamma+\met$. A very similar signal can originate from supersymmetry with a
light gravitino $\tilde{G}$, which is common for example in gauge mediated
supersymmetry breaking (GMSB). In this case, the lightest neutralino
$\tilde{\chi}_0^1$, if it is the NLSP, always  decays to $\tilde{G}$ and a
photon. Assuming ignorance about the mass spectrum of the two models, 6DSM and
GMSB, the only difference are the spins of the particles involved.

In the following we study how GMSB and 6DSM can be distinguished in the
$2\gamma+\met$ signal by analyzing the decay distributions of the photons. In
figure \ref{fig:photons}, we show the energy and angular distribution of the
photons, after applying suitable cuts ($E_\gamma > 5$ GeV and
$|\cos\theta_\gamma | < 0.9 $) to ensure that the photons are visible in the
detector. As shown in figure \ref{fig:bosons}, the signal cross section for
$B^\one$-pair production is of the order of $\sim$100 pb up to
$R^{-1}\lesssim500$ GeV, which is about 5 times larger than the
$\tilde{\chi}^1_0 \tilde{\chi}^1_0$ production cross section for equal masses. 
The efficiency due to the two cuts above is high (about 78\% for $R^{-1}=300$
GeV). The main SM background to $2\gamma+\met$ comes from $\gamma \gamma Z$
with invisible $Z$ decay, whose cross section is  about 16 fb $\times$ 0.2 =
3.2 fb after cuts, and thus already much smaller than the expected signal.
Furthermore, the photon energy distribution in 6DSM and GMSB has distinct
kinematic endpoints, which are given by a formula analogous to
eq.~\eqref{eq:edges}. Unlike this, the energy spectrum of the photon in the SM
background is continuous and peaks at $E_\gamma \sim 0$ GeV or $E_\gamma \sim
500$ GeV. Using this fact together with a proper cut on the invariant mass of
the two photons,  this background becomes completely negligible (smaller by
more than 6 orders of magnitude).

From the right frame of figure \ref{fig:photons} we conclude that the photons
in GMSB tend to be more central while in 6DSM they preferably go in the
forward or backward direction. This is due to the  behavior of the mother
particles when they are produced from $e^+e^-$ collision, {\it i.e.}  vector
bosons ($B_\mu^\one$) prefer the forward or backward direction and neutralinos
($\tilde{\chi}^1_0$) are produced more in the central region. This would allow
for a clear discrimination of the two models based on the spins of the
next-to-lightest new particles.

\FIGURE[t]{ 
\centerline{
\epsfig{file=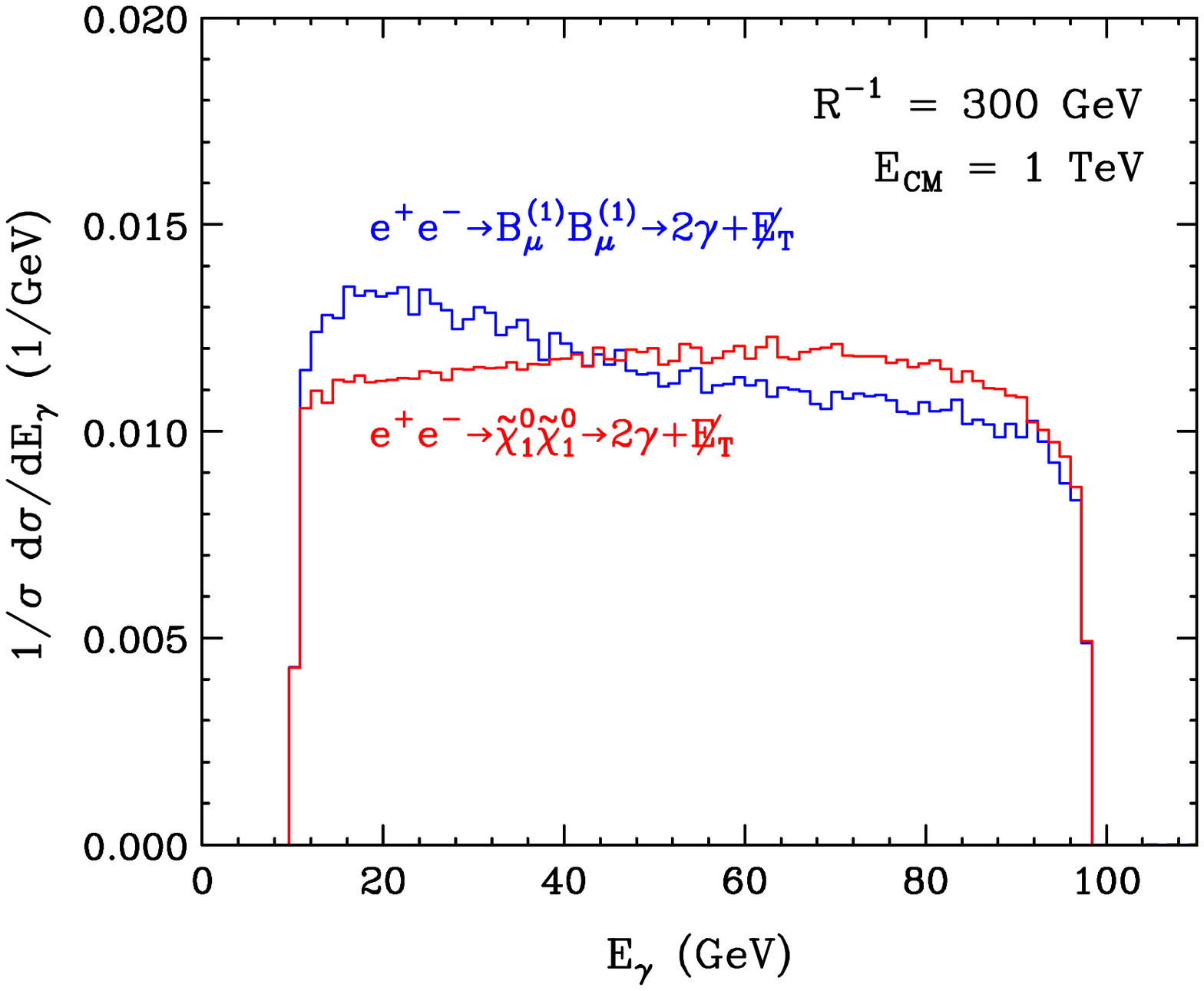,width=7.3cm}
\hspace{0.2cm}
\epsfig{file=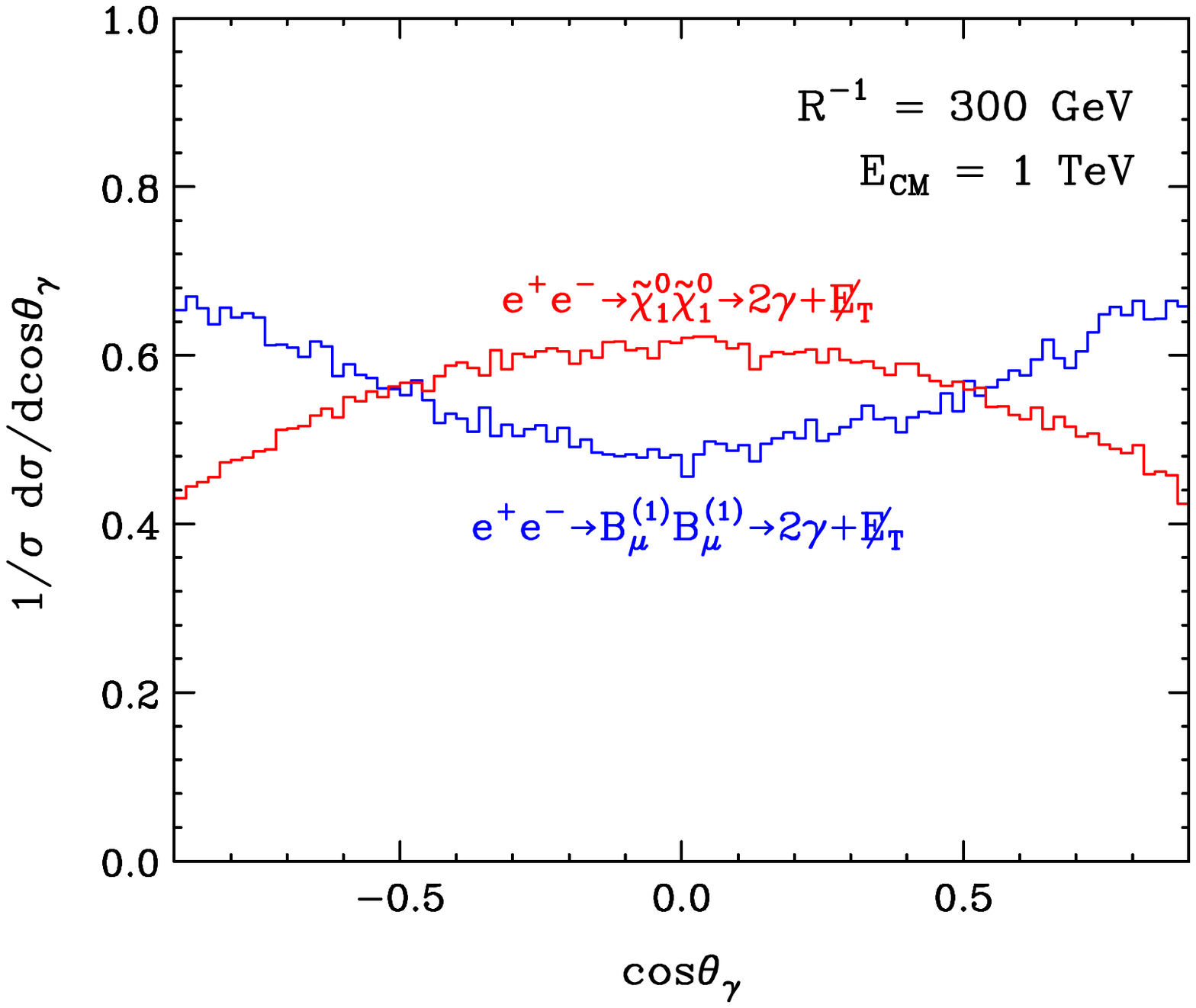,width=7.2cm} 
\caption{\sl (a) Energy and (b) angular distributions of photon in $2\gamma+\met$ 
for 6DSM and supersymmetry with gravitino dark matter} 
\label{fig:photons}}}

Note that the decay $B_\mu^\one \to B^\one_H + \gamma$ is mediated by loop
diagrams, leading to the effective coupling \cite{Dobrescu:2007xf}
\begin{equation}
-\frac{R}{4} \, {\cal C}_{\rm B} \,
\epsilon^{\mu\nu\rho\sigma} F_{\mu\nu} B_{\rho\sigma}^\one B_H^\one, 
\end{equation}
where $F_{\mu\nu}$ and $B^\one_{\rho\sigma}$ are the field strengths of the
photon and the KK hypercharge boson $B_\mu^\one$, respectively. This
interaction reflects the fact that the adjoint scalar $B_H^\one$ is a
pseudo-scalar, with $CP=-1$ quantum numbers. In contrast, 
CP-even scalars would couple to gauge bosons via
\begin{equation}
-\frac{R}{4} \, {\cal C}_{\rm B}' \,
F^{\mu\nu} B_{\mu\nu}^\one B_H^\one. 
\end{equation}
This interaction does not exist in the 6DSM. It would be interesting to explore
if these two cases can be distinguished experimentally by studying the process
$e^+e^- \to B_\mu^\one B_\mu^\one \to \gamma\gamma \, B_H^\one B_H^\one$.
However, an explicit calculation shows that both coupling types lead to the same
squared matrix element, so that a discrimination is not possible. 

This is quite unlike Higgs-strahlung, $e^+e^- \to Z\phi$, where a scalar or
pseudo-scalar Higgs $\phi$ lead to very different angular distributions
\cite{Barger:1993wt}. The crucial discriminative power here comes from the fact
that for the scalar coupling, the longitudinal degree of freedom of the $Z$
boson contributes, while for the pseudo-scalar coupling is does not. In the
decay $B_\mu^\one \to B^\one_H + \gamma$, however, the photon has only
transverse polarization states, so that the distinction between scalar and
pseudo-scalar couplings disappears.

\subsection{Discovery potential for (1,1) KK states}
\label{sec:resonances}

Beyond the lowest (1,0) or, equivalently, (0,1) KK excitations, the 6DSM also
contains many particles with higher KK numbers. Of special interest are states
which have non-zero KK number for each of the two extra dimensions. The
lightest of these are the (1,1) excitations. Since boundary terms break KK
number, and only KK parity is preserved exactly, the (1,1) states can be
produced singly through the interaction of two standard model particles. The
tree-level mass  of the (1,1) particles is $m_{\oo,\rm tree} = \sqrt{2}/R$, so
that they are the lightest states that can be singly produced. In contrast, the
(2,0) excitation have a mass of $m_{(2,0),\rm tree} = 2/R$, which is very
similar to the level-2 states in the 5DSM. Therefore the existence of particles
with even KK parity, but with a mass which is significantly smaller than $2/R$,
is a unique feature of models with more than one universal extra dimension.
This property can be used to distinguish 6DSM and 5DSM, in addition to the
appearance of scalar adjoints.

In $e^+e^-$ collisions, the (1,1) excitations of the neutral gauge bosons can
be produced in the s-channel. In total, there are four (1,1) partners of the
neutral vector bosons: the vector modes $B_\mu^\oo$ and $W_\mu^{\oo 3}$, as
well as the scalar adjoints $B_H^\oo$ and $W_H^{\oo 3}$. The vectors couple to
standard model fermions through the interactions \cite{Burdman:2006gy} 
\begin{equation}
\frac{g \xi^W_f}{16\pi^2} \log \frac{\Lambda^2 R^2}{2} \; 
\bar{f}_{\rm L} \gamma^\mu I^3 f_{\rm L} \, W_\mu^{\oo 3}, 
\qquad
\frac{g' \xi^B_f}{16\pi^2} \log \frac{\Lambda^2 R^2}{2} \; 
\bar{f} \gamma^\mu \frac{Y_f}{2} f \, B_\mu^{\oo},
\label{eq:cplV11}
\end{equation}
where $I^3$ and $Y_f$ are the generators of the third component of weak
$SU(2)_W$ isospin and of hypercharge, respectively, and $g$ and $g'$ are the
couplings of corresponding gauge groups. The $\xi$'s are defined as 
\begin{align}
\xi^W_q &= -4 g_{\rm s}^2 - \frac{1}{2}(\lambda_u^2+\lambda_d^2) + 
\frac{11}{12} g^2 - \frac{1}{12}g'^2,
\displaybreak[0]
\\
\xi^W_l &= - \frac{1}{2}\lambda_l^2 + \frac{11}{12} g^2 -
\frac{3}{4}g'^2,
\displaybreak[0]
\\
\xi^B_{q_{\rm L}} &= -4 g_{\rm s}^2 - \frac{1}{2}(\lambda_u^2+\lambda_d^2) -
\frac{9}{4} g^2 - \frac{83}{12} g'^2,
\displaybreak[0]
\\
\xi^B_{q_{\rm R}} &= -4 g_{\rm s}^2 - \lambda_q^2 - 
\frac{43}{6}  g'^2,
\displaybreak[0]
\\
\xi^B_{l_{\rm L}} &= - \frac{1}{2}\lambda_l^2 - \frac{9}{4} g^2 - 
\frac{91}{12} g'^2,
\displaybreak[0]
\\
\xi^B_{l_{\rm R}} &= - \lambda_l^2 - \frac{59}{6} g'^2.
\end{align}
Here $\lambda_u,\lambda_d,\lambda_l$ are the (generation-dependent) Yukawa
couplings of the up-type quarks, down-type quarks and leptons, respectively, 
which except for $\lambda_t$ can be neglected with good approximation.

The scalars couple to standard model fermions only through higher-dimensional
terms \cite{Burdman:2006gy}:
\begin{equation}
\frac{g \zeta^W_f}{16\pi^2 (\sqrt{2} R^{-1})} \log \frac{\Lambda^2 R^2}{2} \; 
\bar{f}_{\rm L} \gamma^\mu I^3 f_{\rm L} \, D_\mu W_H^{\oo 3}, 
\qquad
\mbox{etc.},
\label{eq:cplH11}
\end{equation}
with $\zeta^W_f \sim {\cal O}(g_{\rm s}^2, g^2)$. By using equations of motion,
it can be seen that couplings of type eq.~(\ref{eq:cplH11}) are suppressed by
the mass of the fermion $f$. Consequently, $B_H^\oo$ and $W_H^{\oo 3}$ cannot be
produced singly in significant rates at an $e^+e^-$ collider.  

In the following we concentrate on the production of (1,1) vector
bosons. Since the interaction terms eq.~(\ref{eq:cplV11}) are suppressed by a
loop factor, the widths of the (1,1) vectors in quite small. For example, for
$R^{-1} = 300$ GeV one obtains at one-loop level:
\begin{align}
M_{B_\mu^{\oo}}&= 415 \text{ GeV},
&
\Gamma_{B_\mu^{\oo}}&= 0.088  \text{ GeV} 
\approx 2.1 \times 10^{-4} M_{B_\mu^{\oo}},
\\
M_{W_\mu^{{\oo}3}}&= 465 \text{ GeV},
&
\Gamma_{W_\mu^{\oo3}}&= 0.20  \text{ GeV} 
\approx 4.4 \times 10^{-4} M_{W_\mu^{\oo3}}.
\end{align}
Such narrow resonances in the s-channel are difficult to observe in $e^+e^-$
collisions, unless by coincidence the center-of-mass energy is close to the mass
of the resonances, $\sqrt{s} \approx M_{B_\mu^{\oo}}, M_{W_\mu^{\oo3}}$. 
If $M_{B_\mu^{\oo}}, M_{W_\mu^{{\oo}3}} < \sqrt{s}$, however, these particles can
be produced via radiation of an additional photon,
\begin{equation}
e^+e^- \to V_\mu^{\oo} + \gamma,  \label{eq:rad}
\end{equation}
where $V = B,W^3$. 
\FIGURE[t]{
\centerline{\epsfig{file=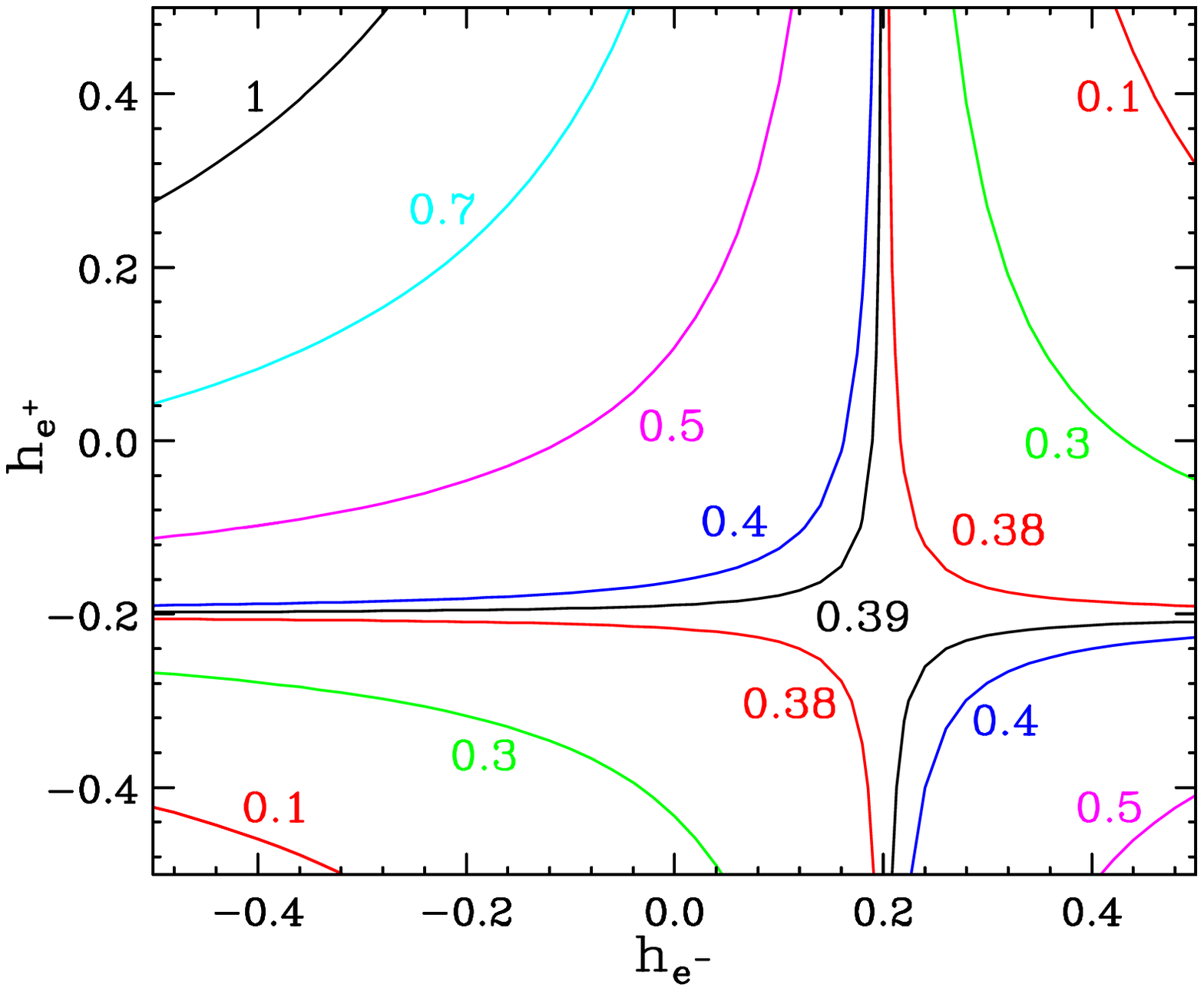,width=7.4cm}
\caption{\sl Cross section for $e^+e^- \to \gamma B_\mu^{(1,1)}$ in helicity basis 
with $E_\gamma > 5$ GeV
and $| \cos\theta_\gamma | < 0.9$.}
\label{fig:xsectionBH11}}}
Figure \ref{fig:xsectionBH11} shows the cross section (in fb) for $e^+e^- \to B_\mu^{\oo} + \gamma$ 
in the helicity bases of the $e^+e^-$ beams.
Since initial-state radiation is dominated by collinear
emission, the photon will be mostly lost unobserved in the beam pipe. Therefore
one has to focus on the final state decay products of the $V_\mu^{\oo}$ in
order to discriminate it from the standard model background. Both $B_\mu^\oo$ and
$W_\mu^{\oo 3}$ have very small couplings to standard model leptons, leading to
suppressed branching fractions BR$[B_\mu^\oo \to e^+e^-] \approx 1$\% and
BR$[W_\mu^{\oo 3} \to e^+e^-] \approx 0.015$\%. Thus a discovery in the leptonic
final state channels does not appear promising. On the other hand, there are
large branching fractions into jets (including all quark flavors except the top
quark): BR$[B_\mu^\oo \to jj] \approx 70$\% and BR$[W_\mu^{\oo 3} \to jj] \approx
70$\%. Therefore, in spite of the limited precision for jet energy measurements,
the di-jet final state is the most promising channel for discovering the (1,1)
vector bosons.

The capability for finding a narrow vector resonance through the radiative
return process eq.~(\ref{eq:rad}) has been analyzed in detail in
Ref.~\cite{Freitas:2004hq}, including standard model backgrounds and detector
resolution effects. Several run scenarios for an $e^+e^-$ linear collider have
been considered: {\it (1)} a scan at the $WW$ pair production threshold
$\sqrt{s} \approx 170$ GeV with 50 fb$^{-1}$ integrated luminosity, {\it (2)} a
scan at the $tt$ pair production threshold $\sqrt{s} \approx 350$ GeV with 100
fb$^{-1}$, {\it (3)} a first-stage high-energy run at  $\sqrt{s} \approx 500$
GeV  with 500 fb$^{-1}$, {\it (4)} and a second-stage run at  $\sqrt{s} \approx
1000$ GeV  with 1000 fb$^{-1}$. The estimated reach versus the predicted
cross section is shown in figure~\ref{fig:11state}. The figure has been normalized
to the case of a vector boson with the same couplings as the standard model $Z$
boson, however this normalization does not affect the conclusions.

As evident from the figure, a 1 TeV $e^+e^-$ collider can discover the
$B_\mu^\oo$ boson for any mass $M_{B_\mu^{\oo}} < 1$ TeV. In addition, a
measurement of the di-jet invariant mass will provide a (moderately precise)
determination of the $B_\mu^\oo$ boson mass.
On the other hand, the $W_\mu^{\oo3}$ has a much smaller cross section owing to
its smaller coupling to $e^+e^-$, so that a direct discovery of this particle
would only be possible if the center-of-mass energy is close to its mass.

However, once a signal for the $B_\mu^\oo$ is detected, one could obtain more
information by tuning the collider energy to its mass. Moreover, one could look
for the $W_\mu^{\oo3}$ by setting the center-of-mass energy closely above
$\sqrt{s} \approx 1.1 \times M_{B_\mu^{\oo}}$, where the $W_\mu^{\oo3}$
resonance is expected in the minimal 6DSM model.

\FIGURE[t]{ 
\epsfig{file=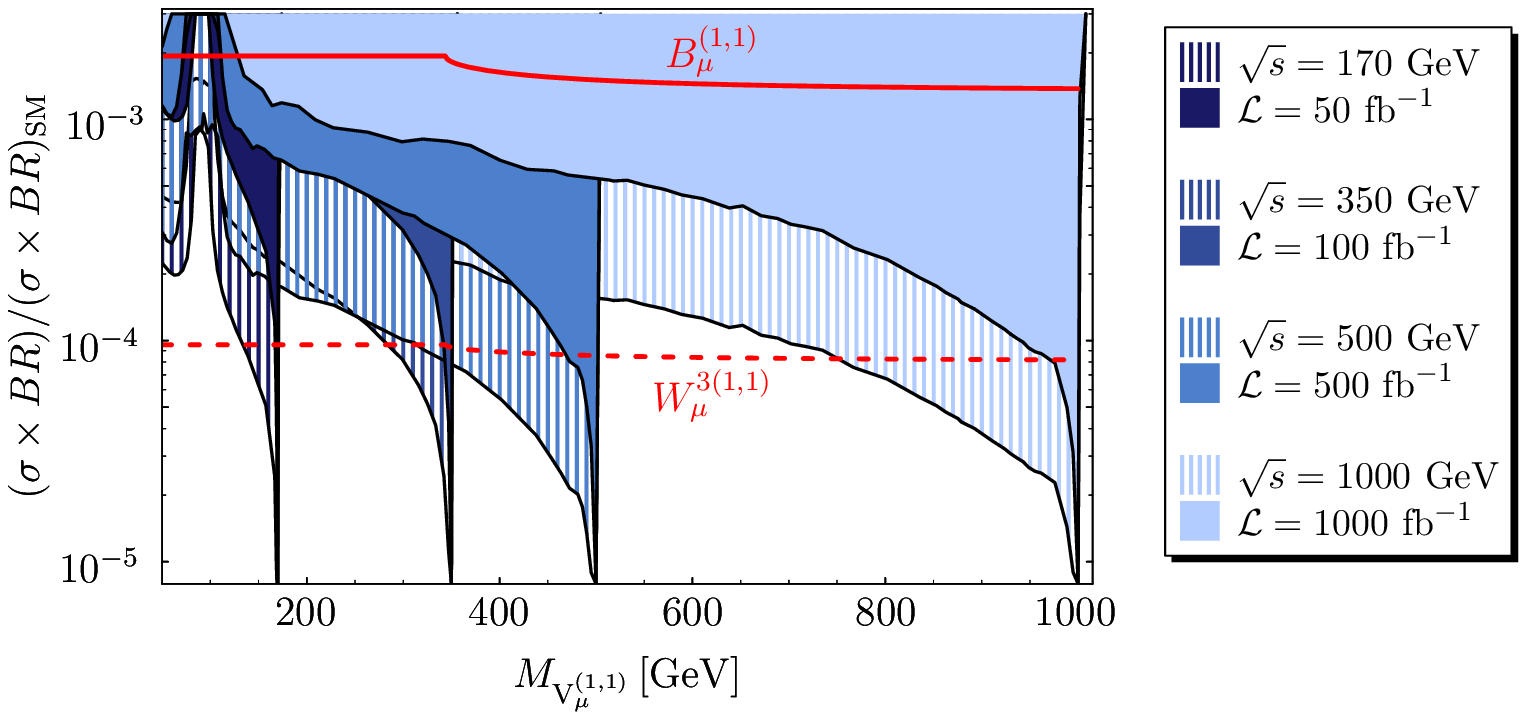, width=14cm, bb=80 460 529 660}
\caption{\sl Projected sensitivity of ILC of $B_\mu^\oo$ and 
$W_\mu^{\oo 3}$ for different run scenarios, as function of the (1,1) vector
boson mass. The hatched regions correspond to
the 95\% confidence level exclusion reach, while the solid regions indicate the
5$\sigma$ discovery reach.}
\label{fig:11state}}
By tuning the center-of-mass energy on the s-channel resonances of $B_\mu^\oo$ 
and $W_\mu^{\oo3}$, the masses, branching fractions and couplings ratios of
these particles can be determined to the per-cent level with only a few
fb$^{-1}$ luminosity \cite{Freitas:2004hq}. However, it will be very difficult
to resolve the non-zero widths of these resonances, since they are comparable
to the expected intrinsic beam energy spread of order ${\cal O}(10^{-4})$, and
are further washed out by beam- and bremsstrahlung. A measurement of the total
width of the (1,1) vector bosons would be very interesting, because it would
allow to determine the absolute coupling strength and thus obtain an estimate
of the scale $\Lambda$ of the UV completion\footnote{Admittedly, such an
estimate of $\Lambda$ would be highly model-dependent since non-perturbative
boundary terms could modify eq.~\eqref{eq:cplV11}}. In principle, a 
determination of the absolute coupling strength of (1,1) vector bosons is
feasible by measuring the invisible decay modes $B_\mu^\oo,W_\mu^{\oo3} \to
\nu\bar{\nu}$ with the help of an initial-state photon for tagging
\cite{Freitas:2004hq}. However, the photon-tagging requirement strongly reduces
the statistics for this kind of measurement, so that it would only be possible
to establish an upper bound on the branching fractions of $B_\mu^\oo$  and
$W_\mu^{\oo3}$ into neutrinos. Nevertheless, this would already allow a rough
determination of the absolute coupling strength to the level of about 6\% for
the $B_\mu^\oo$ (This estimate was derived from the analysis in
Ref.~\cite{Freitas:2004hq}).

\vspace{\medskipamount}

\FIGURE[t]{ 
\centerline{\epsfig{file=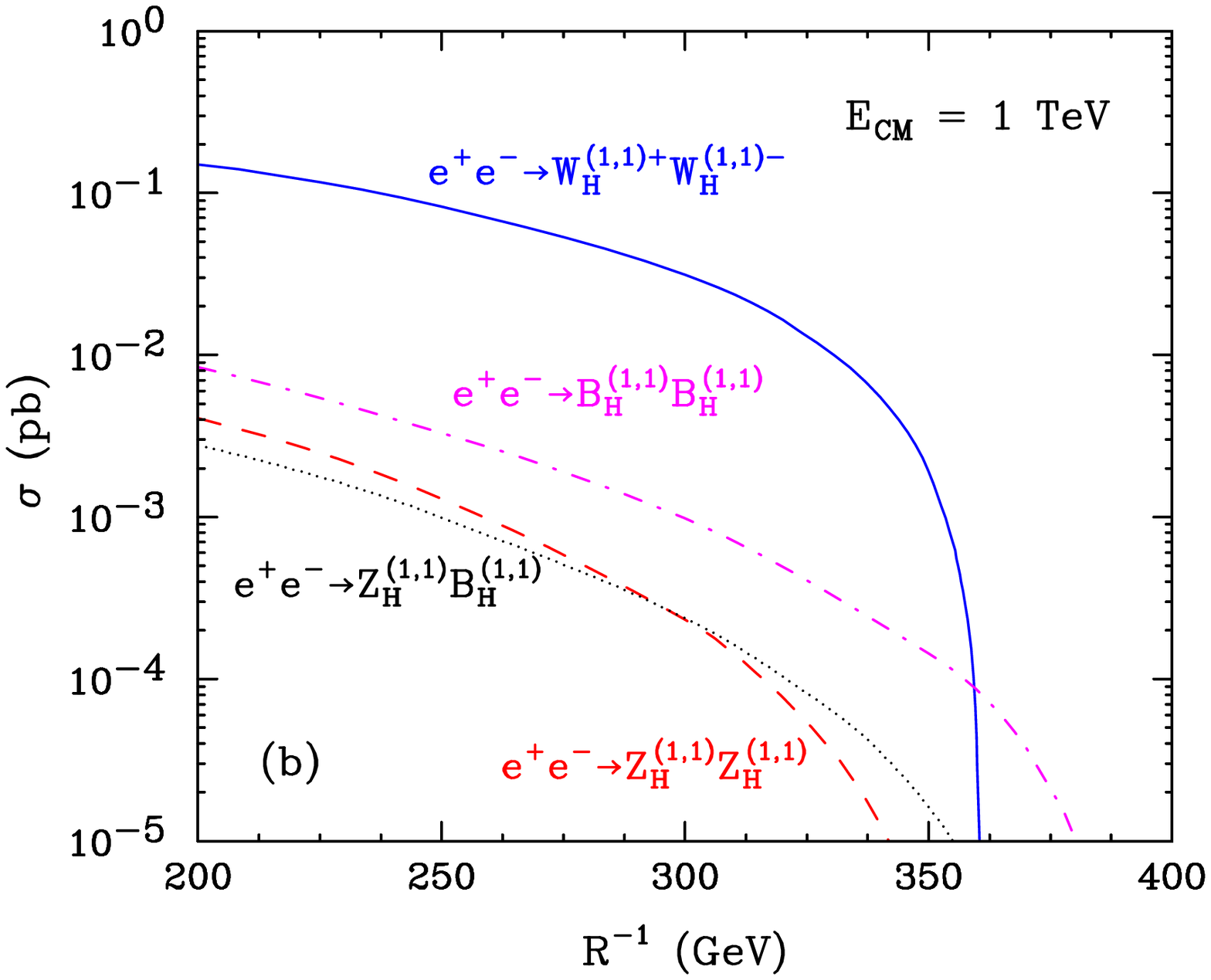,width=7.8cm}
\caption{\sl Pair production cross section for spinless adjoints of KK-level
(1,1).} 
\label{fig:xsection11}}}
While, as pointed out above, the single production of the (1,1) spinless
adjoints in $e^+e^-$ collisons is highly suppressed, pair production of of these
particles is not suppressed by the KK-number violating interaction. Their
production mechanisms are exactly the same as for (1,0) pair production,  as
discussed in section \ref{sec:prod_bosons}, and the cross sections are large for
low $R^{-1}$, as shown in figure \ref{fig:xsection11}. Once these two spinless
adjoints are produced, they decay into the heaviest standard model fermion
pair,  if kinematically allowed. For instance, $Z_H^{(1,1)}$ and $B_H^{(1,1)}$
decay into $t\bar{t}$,  and $W_H^{(1,1)+}$ decays into $t \bar{b}$. 

For pair production of $Z_H^{(1,1)}$ or $B_H^{(1,1)}$ (or mixed $Z_H^{(1,1)}
B_H^{(1,1)}$ pairs),  $t\bar{t}t\bar{t} \to b\bar{b}b\bar{b} W^+W^-W^+W^-$ is
expected in the final state. Up to $R^{-1} \sim 300$ GeV, from figure
\ref{fig:xsection11},  at least a few tens of such events can be observed with a
luminosity of 500 fb$^{-1}$, which already can be sufficient for a discovery
since the standard model background is extremely small.  An interesting
signature might be $\ell^{\pm}\ell^\pm + N~{\rm jets} + \met + X$ with $N\geq4$ 
and $\ell$ being $e$ or $\mu$. In order to reconstruct the mass of the
$Z_H^{(1,1)}$ or $B_H^{(1,1)}$ from the decay products, the fully hadronic decay
of two W's needs to be considered. The separate mass reconstruction of
$Z_H^{(1,1)}$ and $B_H^{(1,1)}$, however, will be difficult due to 
combinatorial background from many jets and the fact that these two spinless
adjoints are close in mass. Nevertheless a rough estimate of the overall mass
scale can be made. This estimate could then be made more precise by tuning the
center-of-mass energy near the threshold, allowing a separation of the two mass
states. From a measurement of the $\beta^3$ behavior ($\beta =  \sqrt{1- {4
m_{(1,1)}^2}/{s}}$) of the cross section near threshold it could be confirmed
that the produced particles are scalar.

For $W_H^{(1,1)+} W_H^{(1,1)-}$ production the signature is  $t \bar{t} b
\bar{b} \to b\bar{b}b\bar{b} W^+W^-$, with the dominant background being
$t\bar{t}$ with two jets from final state radiation, which is expected to be
small. Furthermore, the cross section for $W_H^{(1,1)+} W_H^{(1,1)-}$ is
approximately one order of magnitude larger than the production cross section of
the neutral scalars due to $\gamma$ and $Z$ exchange in the $s$-channel. An
accurate mass measurement of $W_H^{(1,1)^\pm}$ could provide evidence that
$W_H^{(1,1)^\pm}$ and  $Z_H^{(1,1)}$ form an $SU(2)_W$ triplet.

Unfortunately the reach
of a 1~TeV ILC for the electroweak spinless adjoints is limited to $R^{-1}
\lesssim 350$ GeV due to the heaviness of the (1,1) states.
Note that the final state signature for the electroweak spinless adjoints is
quite similar to the (1,1) spinless adjoint of the gluon, which could be
observable at hadron colliders \cite{Dobrescu:2007yp}. 

\section{Conclusions}
\label{sec:conclusions}

Universal extra dimensions are an attractive candidate for new physics, which
would explain the naturalness of the electroweak scale. In contrast to the
extension of the standard model by one extra dimension (5DSM), the standard
model with two extra dimensions (6DSM) has elementary gauge bosons with six
components, one of which becomes an independent scalar particle degree of
freedom. These kind of particles are usually referred to as scalar adjoints. The
first Kaluza-Klein excitation of the scalar adjoint of the hypercharge gauge
group is expected to be the lightest Kaluza-Klein particles and thus would
constitute a good dark matter candidate. Therefore, the 6DSM has a rich
phenomenology at future colliders, with multi-lepton final states, since the
KK-gauge bosons can decay into the corresponding scalar adjoints. Moreover, the
scalar dark matter particle is different from supersymmetry and the 5DSM, where
the new stable particle typically would be a fermion or vector boson,
respectively. 
Other characteristic features of the 6DSM is the decay of the KK-level-1
hypercharge vector boson into the hypercharge scalar adjoint (which is the
lightest KK particle, LKP) and a photon through a loop mediated interaction, and
the presence of higher KK-states which have non-zero KK numbers for both extra
dimensions. The lowest of the these states are the level (1,1) KK excitation,
with a mass that is only larger by factor a $\sqrt{2}$ compared to the (1,0)
states.

In this work, we have analyzed how these properties could be studied
experimentally at a future $e^+e^-$ collider such as ILC. We have computed the
cross sections for the production of various KK particles at
ILC  with $\sqrt{s}=1$ TeV as a function of the size $R$ of the extra dimensions
and discussed the decay pattern of each particle. The rates for many of the new
particles  are large enough to be observable at the ILC if they are within
kinematical reach.
We found that a clean determination of the LKP spin is possible in the
production of KK-electrons, which decay into standard model electrons and the
LKP.

In order to study the 1-loop decay of the KK hypercharge gauge boson into a
photon, we implemented this interaction into {\tt CalcHEP/CompHEP}, which will be
useful for further studies. We were able to show that the signal for this decay
is very clean at the ILC, and using the angular distribution, can be
distinguished from the decay of neutralinos into gravitinos in gauge-mediated
supersymmetry breaking. We further tried to find a method for determining the CP
properties of the LKP, but found that the process does not have any sensitivity
to the CP quantum numbers.

Finally, we have studied the production of (1,1) vector bosons and spinless
adjoints. The vector bosons can be produced singly in $e^+e^-$ collisions, but
with rather small cross sections. Nevertheless, a discovery of the (1,1)
hypercharge gauge boson is possible at the ILC, if its mass is smaller than the
maximum center-of-mass energy of the collider. For the (1,1) spinless adjoints,
the single-production cross sections are too small for an observation, but if
light enough, they could be pair-produced in sizable rates.

In summary, if TeV-scale universal extra dimensions exist, the predicted KK 
excitations could be discovered at future colliders, leading to a rich and
exciting phenomenology. At a high-energy $e^+e^-$ collider, the spectrum,
quantum numbers and decay patterns of the KK particles can be determined
precisely, thus allowing a window into the underlying structure of the
compactification.

\bigskip

{\bf Acknowledgments:} \ We would like to thank B.~Dobrescu and R.~Mahbubani for
useful discussions and helpful comments on the manuscript. ANL is supported by
the U.S. Department of Energy, Division of High Energy Physics, under Contract
DE-AC02-06CH11357. Fermilab is operated by Fermi Research Alliance, LLC under
Contract No. DE-AC02-07CH11359 with the U.S. Department of Energy.  During the
early stages of the project, AF was supported by the Swiss Nationalfonds, and is
grateful to the University of Zurich for the pleasant working environment during
that time.


 \vfil \end{document}